\documentclass[aps,prb,twocolumn]{revtex4-1}

\usepackage{graphicx}
\usepackage{bm}
\usepackage{amssymb}

\begin{document}

\title{Photonic heat transport across a Josephson junction}
\author{George Thomas}
\affiliation{QTF Centre of Excellence, Department of Applied Physics, Aalto University, P.O. Box 15100, FI-00076 Aalto, Finland}

\author{Jukka P. Pekola}
\affiliation{QTF Centre of Excellence, Department of Applied Physics, Aalto University, P.O. Box 15100, FI-00076 Aalto, Finland}

\author{Dmitry S. Golubev}
\affiliation{QTF Centre of Excellence, Department of Applied Physics, Aalto University, P.O. Box 15100, FI-00076 Aalto, Finland}

\begin{abstract}
We present a detailed study of photonic heat transport across a Josephson junction coupled to two arbitrary linear circuits having different temperatures.
First, we consider the linear approximation, in which a nonlinear Josephson potential is replaced by a quadratic one and
the junction acts as an inductor. Afterwards, we discuss the effects of junction anharmonicity. 
We separately consider the weak-coupling limit, in which the Bloch band structure of the junction energy spectrum
plays an important role, and the opposite strong-coupling regime.
We apply our general results to two specific models: a Josephson junction coupled to  two Ohmic resistors and  two resonators.
We derive simple analytical approximations for the photonic heat flux in many limiting cases. 
We demonstrate that electric circuits with embedded Josephson junctions provide a useful platform for quantum thermodynamics experiments. 
\end{abstract}

\maketitle

\section{Introduction}

Transport of heat in nanostructures is a subject of intense research \cite{Giazotto1,Hanggi,Dubi,Sai2016}. 
In most nanoscaled devices, the heat is transferred by either electrons, phonons, or photons. 
Photonic heat transport mechanism often dominates over the two other mechanisms \cite{Cleland} at  temperatures below 100 mK.
In addition, it provides convenient way of transmitting tiny amounts of heat over macroscopic distances \cite{Timofeev,Matti}.
Hence, proper understanding of photonic heat transport is essential, for instance, for development and calibration
of highly sensitive low temperature radiation detectors. 

\begin{figure}
\includegraphics[width=0.9\columnwidth]{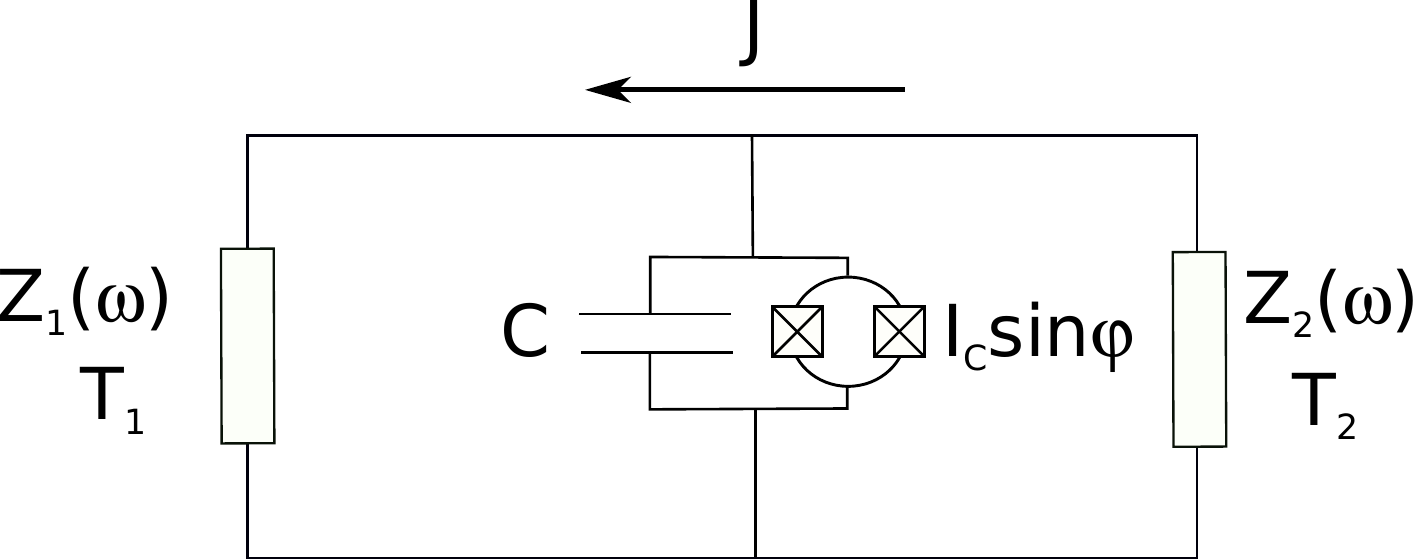}
\caption{SQUID with the critical current $I_C$, tunable by magnetic flux,  shunted by the capacitor $C$ and coupled to the two linear circuits. 
The latter are characterized by the impedances $Z_1(\omega)$ and $Z_2(\omega)$ and temperatures $T_1$ and $T_2$.
The heat flux $J$ flows from the circuit 2 into the circuit 1. It can be controlled by the magnetic flux changing  the critical current $I_C$.}
\label{schematics}
\end{figure}

Photonic heat flux can be accurately controlled by a tunable element embedded in the electric circuit.
The natural choice of such an element for low-temperature superconducting circuits is a SQUID
(superconducting quantum interference device) loop with the Josephson 
critical current adjusted by magnetic flux\cite{Matthias,Alberto}. 
Such systems are promising platforms for realizing quantum thermal machines \cite{Sai2016,Hofer2016}. Motivated by these considerations, 
in this work we theoretically study photonic heat transport through a system
schematically shown in Fig. \ref{schematics}. It contains two linear circuits,
playing the role of thermal baths,  with the impedances $Z_1(\omega)$ and $Z_2(\omega)$,
which include dissipative elements having the temperatures $T_1$ and $T_2$, respectively. These circuits are connected via  
a symmetric SQUID with the critical current modulated by magnetic flux $\Phi$,
\begin{eqnarray}
I_C=I_C(0){\left| \cos({\pi\Phi}/{\Phi_0}) \right|}
\label{IC_Phi}
\end{eqnarray}  
where $\Phi_0$ is the flux quantum. 
The setup of Fig. \ref{schematics} is analogous to the one used in the experiment \cite{Alberto}, in which a SQUID
has been coupled to two resonators terminated by Ohmic resistors. 
It also resembles the usual setup of circuit quantum electrodynamics experiments with transmon qubits\cite{Koch},
which is suitable for heat transport experiments with quantum systems.
For example, the transition from quantum to classical behavior in a Josephson junction
subject to thermal radiation with increasing temperature has been observed in a similar setup\cite{Wallraff}.

The system depicted in Fig. \ref{schematics} also provides
a useful test ground for theoretical predictions about quantum heat transport. 
Indeed, the SQUID is an example of a quantum system coupled to two thermal baths having different temperatures.
It is described by a Hamiltonian of a particle moving in a one-dimensional periodic potential
with the energy spectrum given by a  set of Bloch bands. The Bloch band structure is very sensitive to the ratio
between the Josephson energy of the SQUID $E_J$ and its charging energy $E_C$.  
Depending on the ratio $E_J/E_C$ and the temperatures $T_1$ and $T_2$, 
the model can be approximately reduced to that of weakly scattered particle,
quantum oscillator or two level system\cite{Ojanen,Saito,Ren_2,Segal,Yang,Saito2} coupled to the two thermal baths.
Important related topics are  electronic cooling and
heat rectification in nanostructures\cite{Giazotto1,SN,Hanggi}.
For example, recently a cooler based on a voltage biased Josephson junction has been proposed\cite{Ankerhold_1} and 
heat rectification in nonlinear Josephson circuits has been discussed\cite{Ankerhold_2}.
Thus the physics of the system shown in Fig. \ref{schematics} is very rich.

Approximately replacing the nonlinear Josephson junction by a linear inductor, one can express
the photonic heat flux flowing from the circuit 2 into the circuit 1 in the usual form resembling Landauer formula,
\begin{eqnarray}
J = \int_0^\infty\frac{d\omega}{2\pi} \,\hbar\omega\,\tau(\omega)\,\left[N_2(\omega)-N_1(\omega)\right].
\label{QJ0}
\end{eqnarray}
Here, $N_{1,2}(\omega)=1/(e^{\hbar\omega/k_BT_{1,2}}-1)$ are Bose functions and
$\tau(\omega)$ is the photon transmission probability. 
It equals to the squared absolute value of the microwave transmission coefficient between the ports 1 and 2, 
$\tau(\omega)=|S_{12}(\omega)|^2$, which can be  independently measured in the experiment.  
In Eq. (\ref{QJ0}), we use the sign convention $J>0$ for $T_2>T_1$.
The transmission probability $\tau(\omega)$ can be found in a standard way
by combining microwave impedances of various circuit elements \cite{Hekking}, and in the linear approximation it does not depend on the temperatures 
$T_1$ and $T_2$.

Since the SQUID is a nonlinear, or anharmonic, quantum system, the Landauer formula (\ref{QJ0}) for the heat flux is just an approximation.
There exist several methods of including the nonlinearity into the model. 
In the weak-coupling limit, one can describe the SQUID dynamics by rate equations, accounting for the jumps between its quantum states,
and obtain the heat flux by counting the number of photons emitted to or absorbed from the thermal baths 1 and 2.
Thanks to its simplicity, this method is very popular. It has been used, for example, in the analysis of experimental results 
of Ref. \cite{Alberto}, in  the theoretical description of the heat transport through a superconducting microwave cavity \cite{Tan},
of thermal rectification in a quantum system with discrete energy spectrum \cite{SN}, of Berry phase effects in the heat pumping \cite{Ren}, etc.
More general approach, which is also valid in the strong-coupling limit, has been developed by Ojanen and Heikkil\"a \cite{Ojanen_2}
and Ojanen and  Jauho \cite{Ojanen_1}, who have used the formalism of Keldysh Green's functions, 
similar to the one developed earlier for the description of the electron transport through quantum dots \cite{MW}.
This formalism has been used to develop the theory of thermal rectification in quantum systems \cite{Ojanen_3},
to derive bounds on thermal conductance\cite{Segal_3}, 
and to study heat transport through a two level system \cite{Ojanen,Saito,Ren_2,Segal,Yang,Saito2}. 
An equivalent formally exact approach to the heat transport problems is based on path integral techniques \cite{Tero,QF,Aurell_2,Aurell_3}.
In principle, it allows one to go beyond the average value of the heat flux and to study the full counting statistics of the transferred heat.
Yet another exact and numerically efficient method relies on stochastic Liouville-von Neumann equation with dissipation \cite{Ankerhold_2,Tuorila}.
Interestingly, in many cases one can express the exact heat flux flowing through the nonlinear system 
in the Landauer form (\ref{QJ0}). However, in this case the transmission probability $\tau(\omega)$ becomes dependent
on the bath temperatures $T_{1,2}$.  Such dependence, for example, makes the rectification of heat
possible in a nonlinear and nonsymmetric system, as has been demonstrated in a recent experiment with Josephson junctions \cite{Senior2019}.

In this work, we present a detailed study of heat transport through a Josephson junction, or a SQUID, embedded in the circuit depicted in Fig. \ref{schematics}. 
We use the formalism developed in Refs. \cite{Ojanen_2,Ojanen_1} combining it with the well known results from the quantum theory of Josephson junctions \cite{AZL,Zorin,Averin,SZ}. 
We consider weak and strong coupling regimes, large and small values of the Josephson energy, and various frequency dependencies 
of the environment impedances $Z_{1,2}(\omega)$.
In many limiting cases, we derive analytical expressions both for the photon transmission probability $\tau(\omega)$ and for the heat flux (\ref{QJ0}).  
The paper is organized as follows. In  Sec. \ref{Model}, we introduce the model and the system Hamiltonian; in Sec. \ref{Harmonic}, 
we discuss the linear approximation replacing the junction  by an inductor; in Sec.  \ref{weak_coupling},
we consider the limit of weak coupling; 
in Sec. \ref{strong_coupling}, we analyze the strong-coupling regime,
and in Sec. \ref{conclusion}, we summarize the results. Some details of the calculations and auxiliary information are moved to Appendices.
In Secs. \ref{weak_coupling} and \ref{strong_coupling}, we separately discuss the limits of large, 
$E_J\gg E_C,k_BT_{1,2}$, and small, $E_J\ll E_C,k_BT_{1,2}$, Josephson energy .

\section{Model}
\label{Model}

We consider a symmetric SQUID with the critical current tunable by magnetic flux (\ref{IC_Phi}), 
which we will in the following call junction  for simplicity, 
coupled to  the two linear circuits, characterized by
frequency-dependent impedances $Z_1(\omega)$ and $Z_2(\omega)$, as shown in Fig. \ref{schematics}. The linear circuits form two thermal baths with the temperatures $T_1$ and $T_2$.
The system is described by the Hamiltonian
\begin{eqnarray}
\hat H = \hat H_J + \hat H_1 + \hat H_2 + \hat H_1^{\rm int} + \hat H_2^{\rm int}.
\label{H_full}
\end{eqnarray}
Here
\begin{eqnarray}
\hat H_J = -4E_C\frac{\partial^2}{\partial\varphi^2} + E_J(1-\cos\varphi)
\label{HJ}
\end{eqnarray}
is the Hamiltonian of the junction. 
We model the linear circuits 1 and 2 
as oscillator baths with the Hamiltonians
\begin{eqnarray}
\hat H_j = \sum_k \left[\frac{\hat P_{j,k}^2}{2M_{j,k}} + \frac{M_{j,k}\omega_{j,k}^2\hat X_{j,k}^2}{2}\right], \;\;\; j=1,2,
\end{eqnarray}
where $\hat P_{j,k}$, $\hat X_{j,k}$, $\omega_{j,k}$ and $M_{j,k}$ are, respectively, momentum, coordinate, frequency and mass of
the $k-$th oscillator in the thermal bath $j$.
The interaction Hamiltonians $\hat H_{j}^{\rm int}$ have the form\cite{CL,Grabert}
\begin{eqnarray}
\hat H_j^{\rm int} = -\sum_k \left(c_{j,k} \hat X_{j,k} \hat\varphi - \frac{c_{j,k}^2}{2M_{j,k}\omega_{j,k}^2}\hat\varphi^2 \right), 
\end{eqnarray}
where the coupling constants $c_{j,k}$ should be chosen in such a way that the bath spectral densities are proportional to the real parts of the inverse impedances,
\begin{eqnarray}
\frac{\pi}{2}\sum_k \frac{c_{j,k}^2}{M_{j,k}\omega_{j,k}}\delta(\omega-\omega_k) = \frac{\hbar^2\omega}{4e^2}\,{\rm Re}\left[\frac{1}{Z_j(\omega)}\right].
\label{spectrum}
\end{eqnarray}

Formally exact expression for the heat flux $J$ in terms of Green's functions has been derived in Refs. \cite{Ojanen_2,Ojanen_1}. 
Here we re-derive this expression for our particular setup using perturbation theory. We define the heat flux as
the time derivative of the total energy stored in the environment 1 and write it in the form  
\begin{eqnarray}
J = \frac{d}{dt}\langle \hat H_1 \rangle  = \frac{i}{\hbar} \langle [\hat H_1^{\rm int},\hat H_1] \rangle
= \sum_k  \frac{c_{1,k}}{M_{1,k}}\langle \hat P_{1,k}\hat\varphi\rangle.
\label{Q1}
\end{eqnarray}
The angular brackets here stand for the quantum mechanical averaging with the density matrix of the whole system $\hat\rho$, $\langle\hat A \rangle=\,{\rm tr}\{\hat A\hat\rho\}$. 
The expression (\ref{Q1}) can be further transformed in the interaction representation, in which the operators acquire time dependence 
$\hat A\to \hat A(t)=e^{i\hat H_0t/\hbar}\hat A e^{-i\hat H_0t/\hbar}$ with $\hat H_0= \hat H_J + \hat H_1 + \hat H_2$. 
In this representation the system density matrix satisfies the evolution equation
$i\hbar\, d\hat\rho(t)/dt = [\hat H_{\rm int}(t),\hat \rho(t)]$, where $\hat H_{\rm int}(t)=\hat H_1^{\rm int}(t)+\hat H_2^{\rm int}(t)$. Integrating
this equation over time, we get $\hat\rho(t)=\hat\rho(0)-i\int_0^t dt' [\hat H_{\rm int}(t'),\hat\rho(t')]/\hbar$. Substituting this expression
in  Eq. (\ref{Q1}) and taking the long time limit, we find
\begin{eqnarray}
J = \frac{i}{\hbar}\int_{-\infty}^t dt' \sum_k  \frac{c_{1,k}}{M_{1,k}}\,{\rm tr}\left\{\hat P_{1,k}(t)\hat\varphi(t)[\hat\rho(t'),\hat H_{\rm int}(t')]\right\}
\nonumber\\
= -\frac{i}{\hbar}\int_{-\infty}^t dt' \sum_k  \frac{c^2_{1,k}}{M_{1,k}}\langle [\hat X_{1,k}(t')\hat\varphi(t'),\hat P_{1,k}(t)\hat\varphi(t)]\rangle.
\nonumber
\end{eqnarray}
Note that the term containing $\hat\rho(0)$ vanishes at sufficiently long time $t$, at which the information about the initial state
of the dissipative system is lost. For such times one can also extend the integration
over $t'$ as follows, $\int_0^t dt' \to \int_{-\infty}^t dt'$.
Furthermore, within the framework of perturbation theory one can factorize the averages of the products of four operators
into the products of the pairwise averages $\langle \hat X_{1,k}(t')\hat P_{1,k}(t)\rangle$ and $\langle\hat\varphi(t)\hat\varphi(t') \rangle$.  
The average $\langle \hat X_{1,k}(t')\hat P_{1,k}(t)\rangle$ can be evaluated, since in the lowest order of the perturbation theory, 
the oscillators of the thermal baths do not interact with the junction.
Performing these operations, we arrive at the expression
\begin{eqnarray}
J &=&  
\int\frac{d\omega}{2\pi}  \frac{\hbar^2\omega^2}{4e^2}\,{\rm Re}\left[\frac{1}{Z_1(\omega)}\right]
\nonumber\\ && \times\,
\big(S_{\varphi}(-\omega)[1+ N_1(\omega)]-S_\varphi(\omega)  N_1(\omega) \big).
\label{Q30}
\end{eqnarray}
Here we have introduced the Fourier transformed phase-phase correlation function
\begin{eqnarray}
S_\varphi(\omega) = \int dt e^{i\omega t} \langle\hat\varphi(t)\hat\varphi(0)\rangle.
\label{Phi}
\end{eqnarray}
Although we have used perturbation theory while deriving the formula (\ref{Q30}), it is actually exact. 
One can prove this result by either Keldysh Green's function technique \cite{Ojanen_2,Ojanen_1}, or by 
path integral technique\cite{QF,Aurell_2,Aurell_3}.

One can alternatively express the heat flux in terms of 
the correlation function of the  operators of the charge accumulated in the junction capacitance $\hat Q = -2i e (\partial/\partial\varphi)$,
\begin{eqnarray}
S_Q(\omega) = \int dt e^{i\omega t} \langle\hat Q(t)\hat Q(0)\rangle.
\label{SQ}
\end{eqnarray}
Since the operators $\hat\varphi(t)$ and $\hat Q(t)$ are related by the equation of motion for the phase,
\begin{eqnarray}
\frac{d\hat\varphi}{dt}=\frac{i}{\hbar}[\hat H,\hat\varphi] = \frac{4E_C}{\hbar e}\hat Q,
\label{comm}
\end{eqnarray}
the correlation function (\ref{Phi}) can be written in the form
\begin{eqnarray}
S_\varphi(\omega)=\frac{16E_C^2}{e^2\hbar^2\omega^2}S_Q(\omega).
\end{eqnarray}
Accordingly, the heat flux (\ref{Q30}) can be expressed as
\begin{eqnarray}
J&=&\frac{4E_C^2}{e^4}\int\frac{d\omega}{2\pi}  \,{\rm Re}\left[\frac{1}{Z_1(\omega)}\right] 
\nonumber\\ &&\times\,
\big(S_Q(-\omega)[1+N_1(\omega)]-S_Q(\omega)N_1(\omega)\big).
\label{Q3}
\end{eqnarray}

\begin{figure}
\includegraphics[width=0.9\columnwidth]{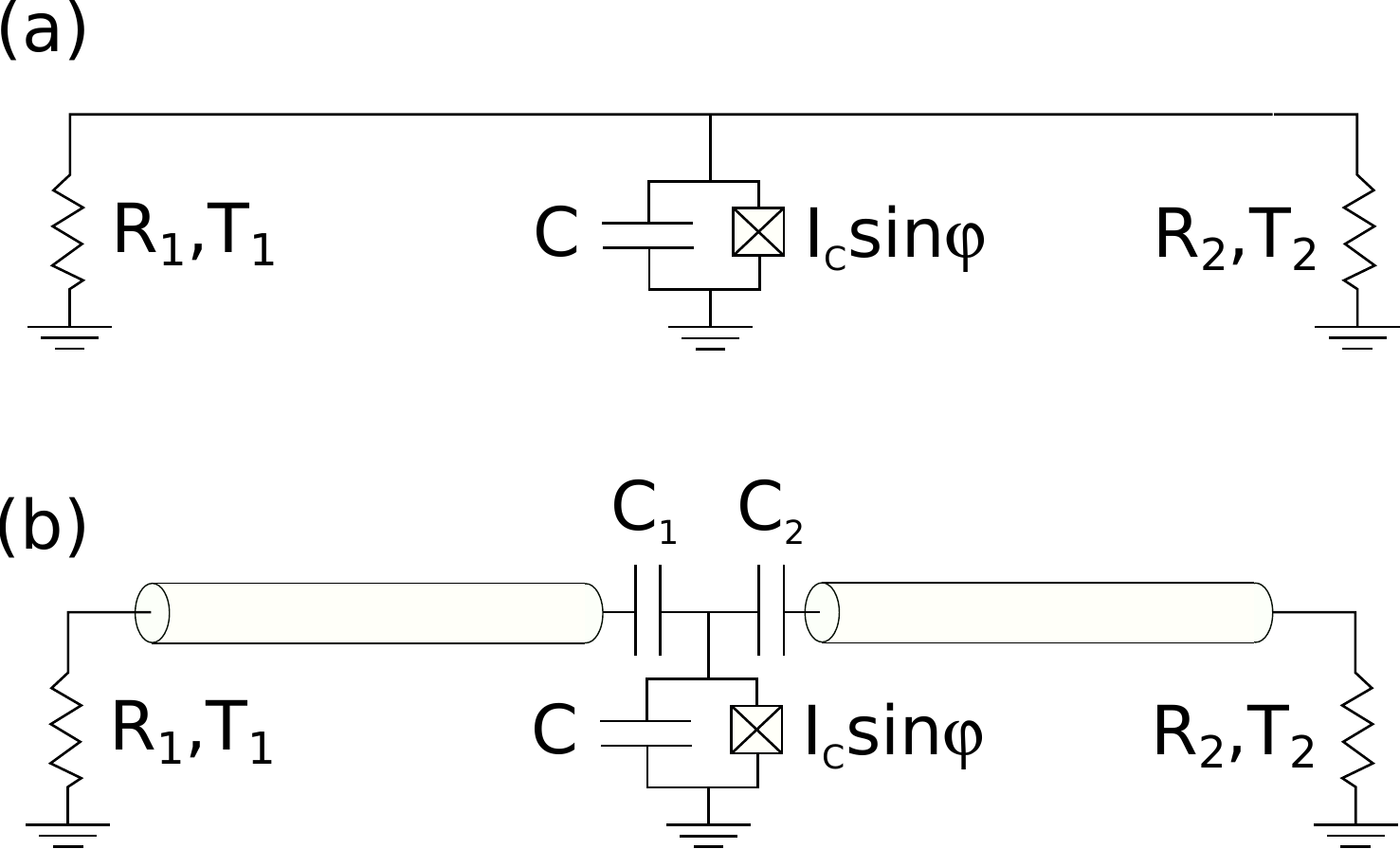}
\caption{Two specific examples: (a) a junction directly coupled to two Ohmic resistors $R_1$ and $R_2$
and (b) a junction coupled to two resonators with the impedances (\ref{Zres}). }
\label{examples}
\end{figure}

Equations (\ref{Q30}) and  (\ref{Q3}) are the starting points of our analysis.
Employing various approximations, we will derive approximate expressions for the heat flux $J$ without specifying 
particular frequency dependence of the impedances $Z_{1,2}(\omega)$. 
Afterwards we will consider two specific examples. One example is the case of Ohmic dissipation
with $Z_j(\omega)=R_j$, which is illustrated in Fig. \ref{examples}a. 
Another  example, shown in Fig. \ref{examples}b and inspired by recent heat transport experiment \cite{Alberto}, 
is the junction capacitively coupled to the two $\lambda/4$ resonators 
having characteristic impedances $Z_{r1}$ and $Z_{r2}$ and terminated by Ohmic resistors $R_1$ and $R_2$.
In this case, the impedances of the two linear electric circuits are
\begin{eqnarray}
Z_j(\omega) &=& \frac{1}{-i\omega C_j} - i Z_{rj}\tan\left[\frac{\pi}{2}\frac{\omega}{\omega_j} + i \alpha_j\right],
\label{Zres}\\
\alpha_j &=& \frac{1}{2}\ln\frac{Z_{rj}+R_j}{Z_{rj}-R_j}.
\label{alpha}
\end{eqnarray}
Here, $C_j$ are the coupling capacitors [see Fig. \ref{examples}(b)], $\omega_j$ are the frequencies of the fundamental modes of the resonators, and the parameters $\alpha_j$ 
determine their internal quality factors, $Q_j=\pi/4\alpha_j$.
Here we will only consider the regime $R_j<Z_{rj}$.
Provided $Q_j\gg 1$, $E_J\gg E_C$ and the temperatures $T_{1,2}$ are sufficiently low, one can keep only the two lowest energy levels formed close to the bottom
of the Josephson potential well, and approximately describe the system by Rabi Hamiltonian \cite{Koch}
\begin{eqnarray}
\hat H = -\frac{\hbar\omega_{01}}{2}\hat\sigma_z 
+ \sum_{j=1,2}\big[ \hbar\omega_j \hat b^\dagger_j \hat b_j+   \hbar g_j (\hat b^\dagger_j + \hat b_j)\hat\sigma_x\big]. 
\label{JC}
\end{eqnarray}
Here the transition frequency between the levels is $\hbar\omega_{01}=\hbar\omega_J-E_C$, $\omega_J=\sqrt{8E_JE_C}/\hbar$ 
is the classical frequency of small oscillations at the bottom of the potential well, 
and the coupling constants $g_1,g_2$ are given by
\begin{eqnarray}
g_j = \sqrt{\frac{Z_{rj}\omega_j C_j^2}{\pi C_\Sigma}}\,\omega_j.
\label{g_j}
\end{eqnarray}
Here we defined the total capacitance $C_\Sigma=C_1+C_2+C$.

\section{Linearized dynamics}
\label{Harmonic}

In this section, we approximately replace the nonlinear Josephson junction by an inductor with the impedance
\begin{eqnarray}
Z_J(\omega)=-i{\hbar\omega}/{2eI_C}.
\label{ZJ0}
\end{eqnarray}
This simple approximation is valid in the wide range of parameters.
For the beginning, it is valid for sufficiently low temperatures, $k_BT_{1,2}\lesssim 2E_J$, sufficiently high ratio $E_J/E_C\gtrsim 1$, 
and sufficiently strong coupling between the junction and the environment,
\begin{eqnarray}
{\rm Re}\left[\frac{R_q}{Z_1(\omega_J)}+\frac{R_q}{Z_2(\omega_J)}\right] \gtrsim 1.
\label{SC}
\end{eqnarray}
Here we have introduced  the resistance quantum $R_q=h/e^2$.
The condition (\ref{SC}) ensures that 
the width of the energy levels formed at the bottom of  the Josephson potential well, 
which is determined by the transition rates (\ref{Gamma_mn}), exceeds the charging energy $E_C$. Since the latter
defines the scale of anharmonicity in the system the junction may be viewed as a linear element.
The linear approximation is also valid at high temperatures $k_BT_{1,2}\gg E_J$, where 
one can just put $I_C=0$,  thus removing the nonlinear element from the circuit.

\begin{figure}
\includegraphics[width=\columnwidth]{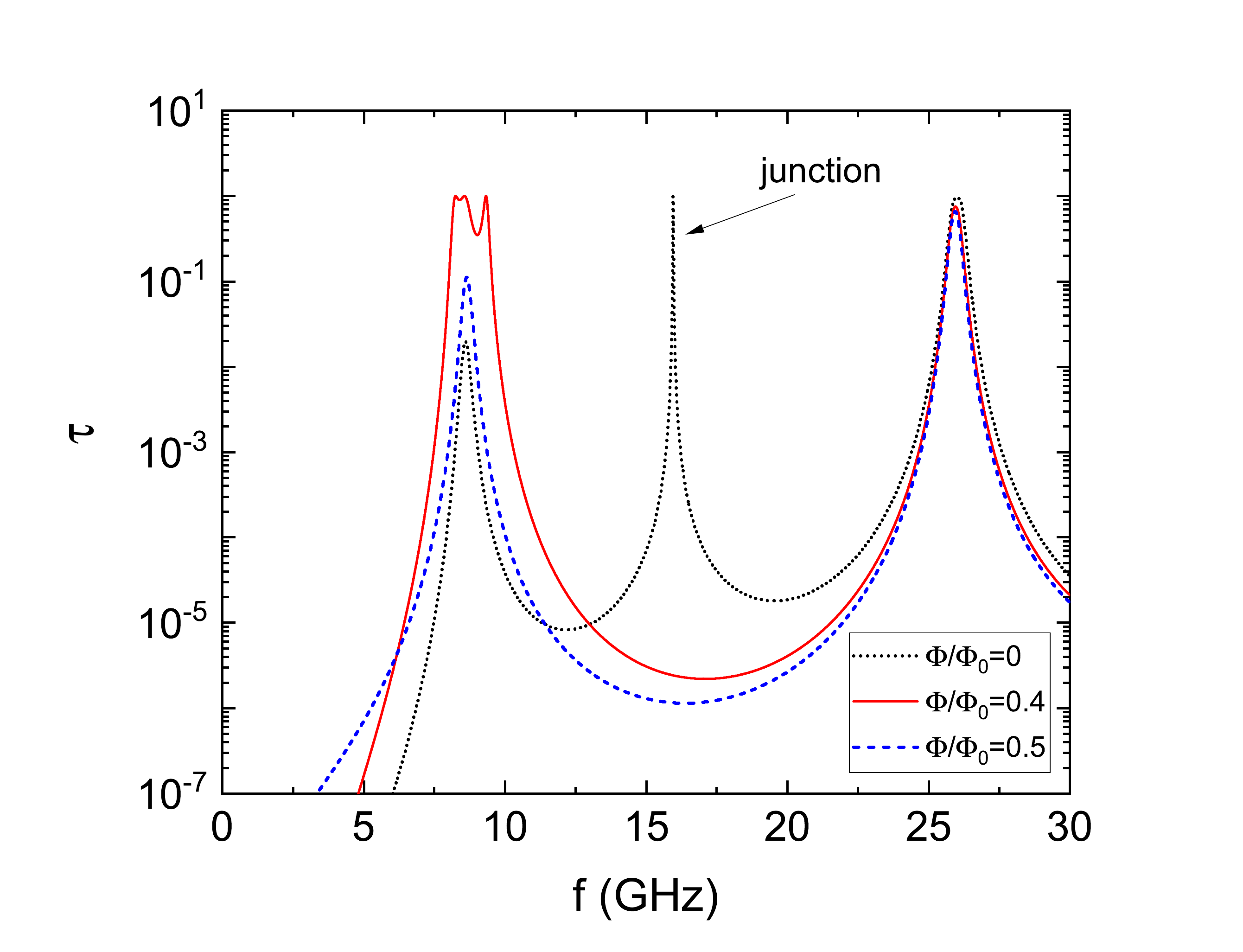}
\caption{Transmission probability $\tau(\omega)$ (\ref{tau}) for a symmetric system of Fig. \ref{examples}b
with two identical resonators and for three different values of magnetic flux.  
The parameters of the system are: $\omega_1/2\pi=\omega_2/2\pi=8.84$ GHz, $Z_{r1}=Z_{r2}=50$ $\Omega$, $R_1=R_2=2$ $\Omega$,  $C_1=C_2=15$ fF, $C=58.7$ fF,
critical current at zero magnetic flux is $I_C(\Phi=0)=291$ nA.
This results in the following values for the charging energy, the Josephson energy and the coupling constants between the junction
and the resonators (\ref{g_j}):  $E_C/h = 218.4$ MHz, $E_J(\Phi=0)/h=144.6$ GHz, and
$g_1/2\pi=g_2/2\pi=418.6$ MHz. 
At $\Phi=0$ the Josephson frequency equals to $\omega_J/2\pi = 15.9$ GHz and it is
far detuned from the resonator frequencies $\omega_{1,2}$. In this case a sharp peak in the transmission probability, indicated by arrow, is formed at frequency $f=\omega_J/2\pi$.
Close to $\Phi/\Phi_0=0.4$ the resonance condition $\omega_J=\omega_{1,2}$ is achieved, the two resonator modes and the junction are hybridized, 
and three peaks in $\tau(\omega)$ are formed. At $\Phi/\Phi_0=0.5$, the critical current is suppressed, $I_C=0$, the peak associated with the junction disappears
and only the peaks coming from the modes of the resonators centered at $\omega_n=(2n+1)\omega_1$ remain.}
\label{Fig:tau_high_EJ}
\end{figure}

Once the Josephson junction has been replaced by a linear lumped element,
one can exactly reduce the full quantum problem to the solution of the classical Langevin equations 
for the Josephson phase and currents, which contain the stochastic noises generated by the circuit elements\cite{Hekking,Schmid}. 
Similar Langevin equations have been used, for example, in order to describe the transport of heat by phonons in harmonic lattices \cite{Dhar}. 
The details of the analysis are given in  Appendix \ref{sec-Langevin}. 
The final expression for the heat flux acquires the Landauer form (\ref{QJ0})
with the photon transmission probability
\begin{eqnarray}
\tau(\omega) = 
\frac{4 {\rm Re}\left[\frac{1}{Z_1(\omega)}\right] {\rm Re}\left[\frac{1}{Z_2(\omega)}\right]}
{\left| -i\omega C + \frac{1}{Z_1(\omega)} + \frac{1}{Z_2(\omega)} +  \frac{1}{Z_J(\omega)} \right|^2}.
\label{tau}
\end{eqnarray}
Here the junction impedance is given by  Eq. (\ref{ZJ0}).
Since $\tau(\omega)$ does not depend on the temperatures of the two baths, the heat flux (\ref{QJ0}) has the property $J(T_1,T_2)=-J(T_2,T_1)$, 
which implies the absence of heat rectification in the linear approximation.
Absence of rectification is the well known property of linear harmonic systems \cite{SN,Dhar1}.

\begin{figure}
\includegraphics[width=\columnwidth]{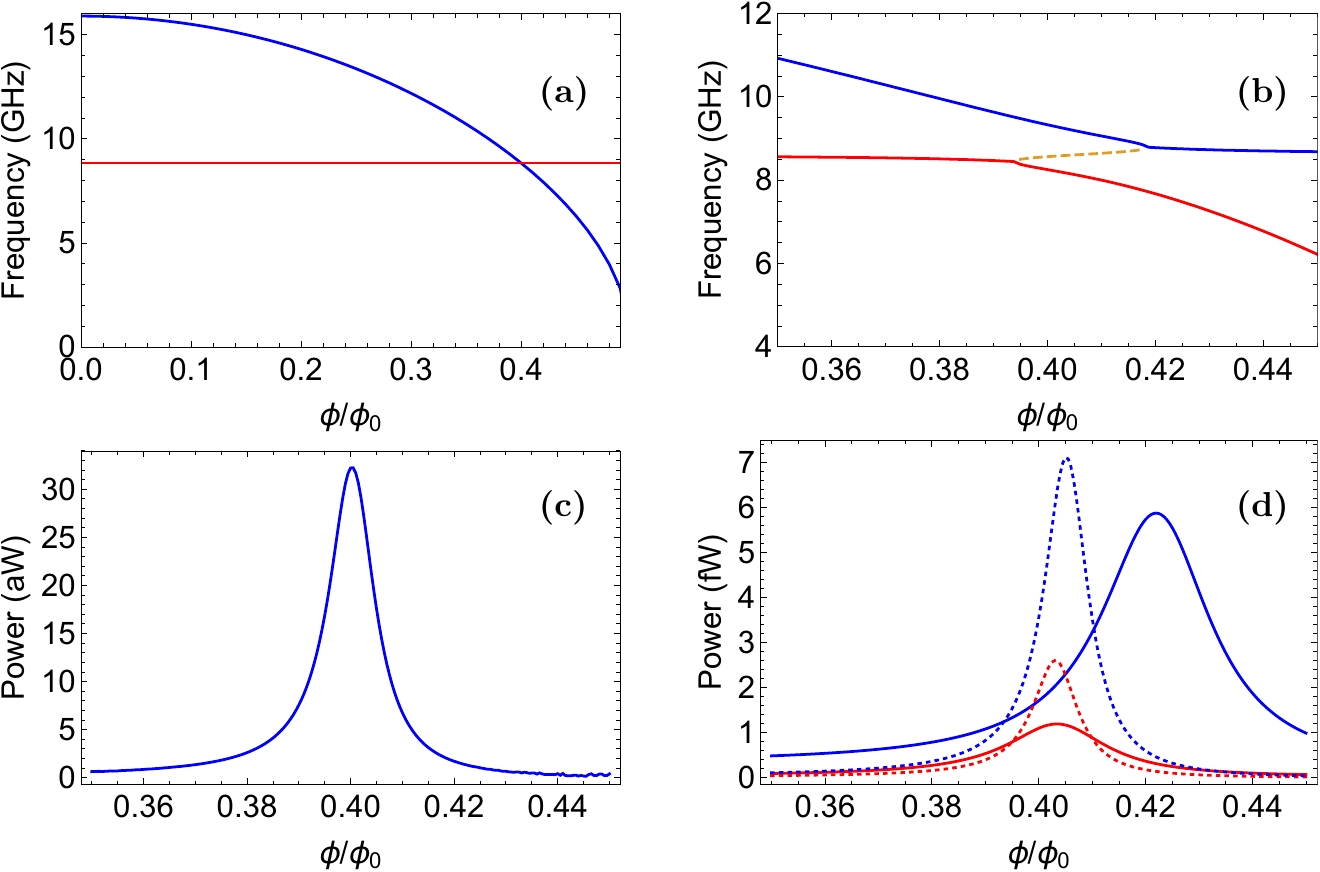}
\caption{ Positions of the peaks in the transmission probability (a,b) and the heat flux (c,d) as a function  of the magnetic flux for
a symmetric system with parameters given in the caption of Fig. \ref{Fig:tau_high_EJ}.
(a) Peak positions in the weak coupling limit with $C_1= C_2=1$ fF.
(b) Peak positions at stronger coupling with $C_1= C_2=15$ fF.  Avoided crossing of the hybridized modes close to the resonance 
and the formation of the third peak, shown by the red dashed line, are clearly visible. 
(c) Heat flux vs magnetic flux in the weak-coupling regime $C_1= C_2=1~ {\rm fF}$. In this case the integral (\ref{QJ0}) 
 and the approximate expression  (\ref{P1_osc_1}) produce the same result.
 (d)  Heat flux vs magnetic flux at stronger coupling 
$C_1= C_2=9~ {\rm fF}$ (red curves)  and  $C_1= C_2=15~ {\rm fF}$ (blue curves).
 The solid curves  are obtained from Eq. (\ref{QJ0}) whereas the dotted curves is from the approximate Eq. (\ref{P1_osc_1}). 
 We have used the following parameters: $f_r=\omega_1/(2\pi)=\omega_2/(2\pi)=8.84$ GHz, $E_J/h=144.6 $ GHz, 
$E_C/h=218.4$ MHz, $T_2=300$ mK, $T_1=150$ mK, $R_1=R_2=2\; \Omega$, and $Z_{r1}=Z_{r2}=50\; \Omega$.
}
\label{anharmonicpart}
\end{figure}

As an example, in Fig. \ref{Fig:tau_high_EJ}, we plot the transmission probability (\ref{tau}) versus frequency $f=\omega/2 \pi$ for a junction coupled 
to the two identical resonators characterized by the impedances (\ref{Zres}). 
We observe that $\tau(\omega)$ has peaks at frequencies corresponding to the eigen-modes of the resonators $\omega_n=(2n+1)\omega_1$, with $n=0,1,2\dots$.
Besides that, there exists a peak at the frequency  of phase oscillations in the junction $\omega_J$.
For a certain value of magnetic flux the resonant condition $\omega_J=\omega_1=\omega_2$
is achieved. Close to this value three peaks appear in $\tau(\omega)$ due to hybridization of the modes of the resonators and of the junction.
The positions of these peaks are shown in Figs. \ref{anharmonicpart}a and \ref{anharmonicpart}b.  For weak coupling (Fig. \ref{anharmonicpart}a) the hybridization of the modes
is almost invisible, while for strong coupling (Fig. \ref{anharmonicpart}b) it is quite strong and manifests itself in avoided crossing of the peak position lines.  
In Figs. \ref{anharmonicpart}c and \ref{anharmonicpart}d we plot the dependence of the heat power (\ref{QJ0}) on magnetic flux for weak and strong coupling respectively.

In certain limiting cases one can derive analytical approximations for $\tau(\omega)$ and for the heat flux $J$.
First, we consider the junction coupled to identical resonators with high quality factors.
We assume that the junction mode is sufficiently far detuned from the modes of the resonators and ignore
the shift of the junction frequency induced by the coupling to the resonators.  
This approximation is valid provided the condition
\begin{eqnarray}
\frac{g_1^2\omega_J^2}{\omega_n^2[(\omega_J-\omega_n)^2+\gamma_1^2]}\ll 1 
\label{detuning}
\end{eqnarray}
is satisfied for all modes of the resonators.
Here $\gamma_1=2\alpha_1\omega_1/(\pi +2Z_{r1}C_1\omega_1)$ is the total damping rate of the resonators. 
In this case, the transmission probability $\tau(\omega)$ is expressed as a sum of well separated peaks,
\begin{eqnarray}
&& \tau(\omega) \approx \frac{\gamma_J^2}{(\omega-\omega_J)^2 + \gamma_J^2} 
\nonumber\\ &&
+\, \sum_{n=0}^\infty \frac{a_n\gamma_1^4}{\left[(\omega-\omega_n)^2+\gamma_1^2-\frac{a_n^2\omega_n^2}{4}\right]^2+a_n\gamma_1^4}. 
\label{tau_app}
\end{eqnarray}
The first Lorentzian peak comes from the junction mode. Its width $\gamma_J$ is given by
\begin{eqnarray}
\gamma_J = \frac{2 g_1^2\omega_J^2\gamma_1}{\omega_1^2[(\omega_J-\omega_{n_0})^2+\gamma_1^2]},
\end{eqnarray}
where $\omega_{n_0}$ is the frequency of the mode closest to the Josephson frequency $\omega_J$. 
The remaining peaks correspond to the modes of the resonators. They have non-Lorentzian shape, and their heights $a_n$ are
determined by the detuning between the junction mode and the corresponding resonator mode,
\begin{eqnarray}
a_n = \frac{4 g_1^2\omega_n^2}{\omega_1^2\gamma_1^2|\omega_n^2-\omega_J^2|}.
\end{eqnarray}
The resonator modes split into the pairs of closely lying peaks if $a_n\omega_n>2\gamma_1$, when
the effective coupling between the modes of the resonators becomes stronger than the dissipation rate.
At high temperatures, $k_B \max\{T_1,T_2\}\gtrsim \min\{\hbar\omega_J,\hbar\omega_1\}$, one can approximately replace all the peaks
in  Eq. (\ref{tau_app}) by $\delta-$functions. The heat flux (\ref{QJ0}) then takes the form
\begin{eqnarray}
J &=& \frac{\gamma_J\hbar\omega_J}{2}[N_2(\omega_J)-N_1(\omega_J)]
\nonumber\\ &&
+\, \sum_{n=0}^\infty \frac{\gamma_1\hbar\omega_n [N_2(\omega_n)-N_1(\omega_n)]}{1+\frac{R_1^2 C_\Sigma^2}{Z_{r1}^4C_1^4\omega_n^2}\left(1-\frac{\omega_J^2}{\omega_n^2}\right)^2}.
\label{P1_delta}
\end{eqnarray} 
The first term in this expression describes the heat transport through the junction mode, while the second term -- through the modes of the resonators. 
The effective coupling between the modes of different resonators depends on the value of the Josephson frequency $\omega_J$. 

\begin{figure}
	\includegraphics[width=0.7\columnwidth]{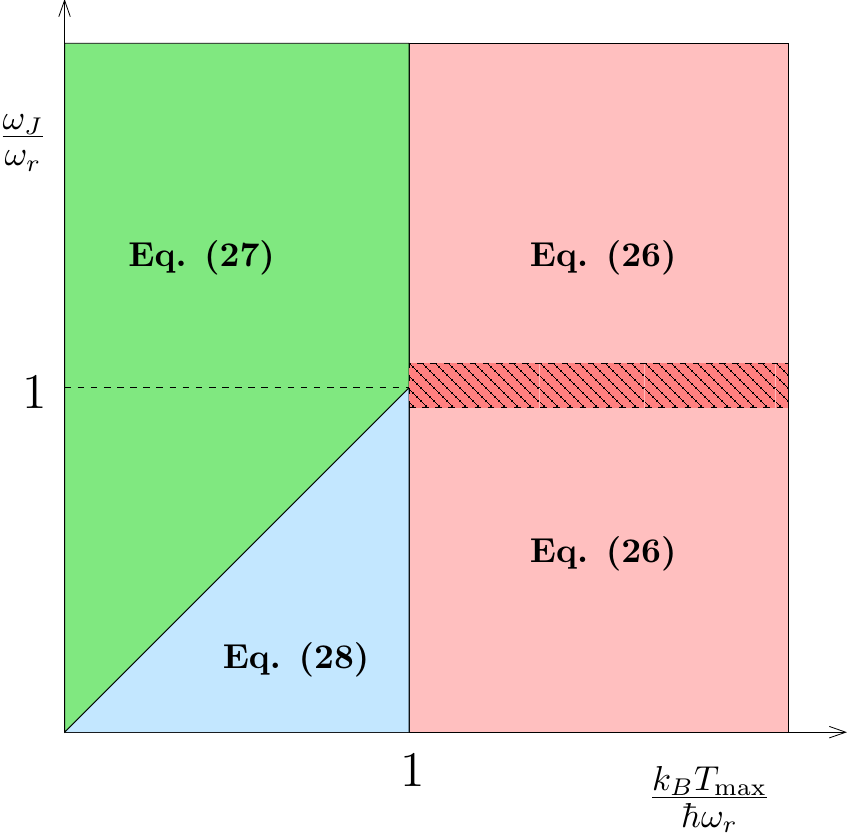}
	\caption{Schematic representation of the heat transport regimes for a Josephson junction weakly coupled to two identical high quality factor resonators. 
	Here $T_{\max}=\max\{T_1,T_2\}$ and $\omega_r=\omega_1=\omega_2$ is the frequency of the resonators. Shaded area
    indicates the region in which $\omega_J\approx \omega_r$ and $k_BT_{\max}\gtrsim \hbar\omega_r$. In this region the
    heat flux approaches maximum values. }
	\label{regimes}
\end{figure}

Next, we consider arbitrary resonators, which are not necessarily identical, but assume that the temperatures are low, 
$k_B \max\{T_1,T_2\} \lesssim \min\{\hbar\omega_J,\hbar\omega_{1,2}\}$. In this case  
the transmission probability and the heat flux are
\begin{eqnarray}
\tau(\omega) &=& \frac{\hbar^4 R_1R_2C_1^2C_2^2}{4e^4E_J^2}\omega^6,\;\; \omega\ll\omega_J,
\nonumber\\
J &=& \frac{\pi^7}{15} \frac{R_1R_2C_1^2C_2^2}{\hbar^3 e^4E_J^2}[(k_BT_2)^8-(k_BT_1)^8].
\label{P1_low_T12}
\end{eqnarray}
In the regime $ \hbar\omega_J \lesssim \hbar\omega, k_B\max\{T_{1},T_{2}\}\ll \hbar\omega_{1,2}$ we find
\begin{eqnarray}
&& \tau(\omega) = \pi\gamma_J\delta(\omega-\omega_J)+ \frac{4R_1R_2C_1^2C_2^2}{C_\Sigma^2}\omega^2, 
\nonumber\\
&& J = \frac{\gamma_Jk_B(T_2-T_1)}{2}+\frac{2\pi^3}{15} \frac{R_1R_2C_1^2C_2^2 k_B^4}{C_\Sigma^2\hbar^3} (T_2^4-T_1^4).
\nonumber\\
\label{Q1_cap}
\end{eqnarray}
In Fig. \ref{regimes} we schematically overview various heat transport regimes
for a Josephson junction coupled to two identical resonators with high quality factors.

\begin{figure}
\includegraphics[width=\columnwidth]{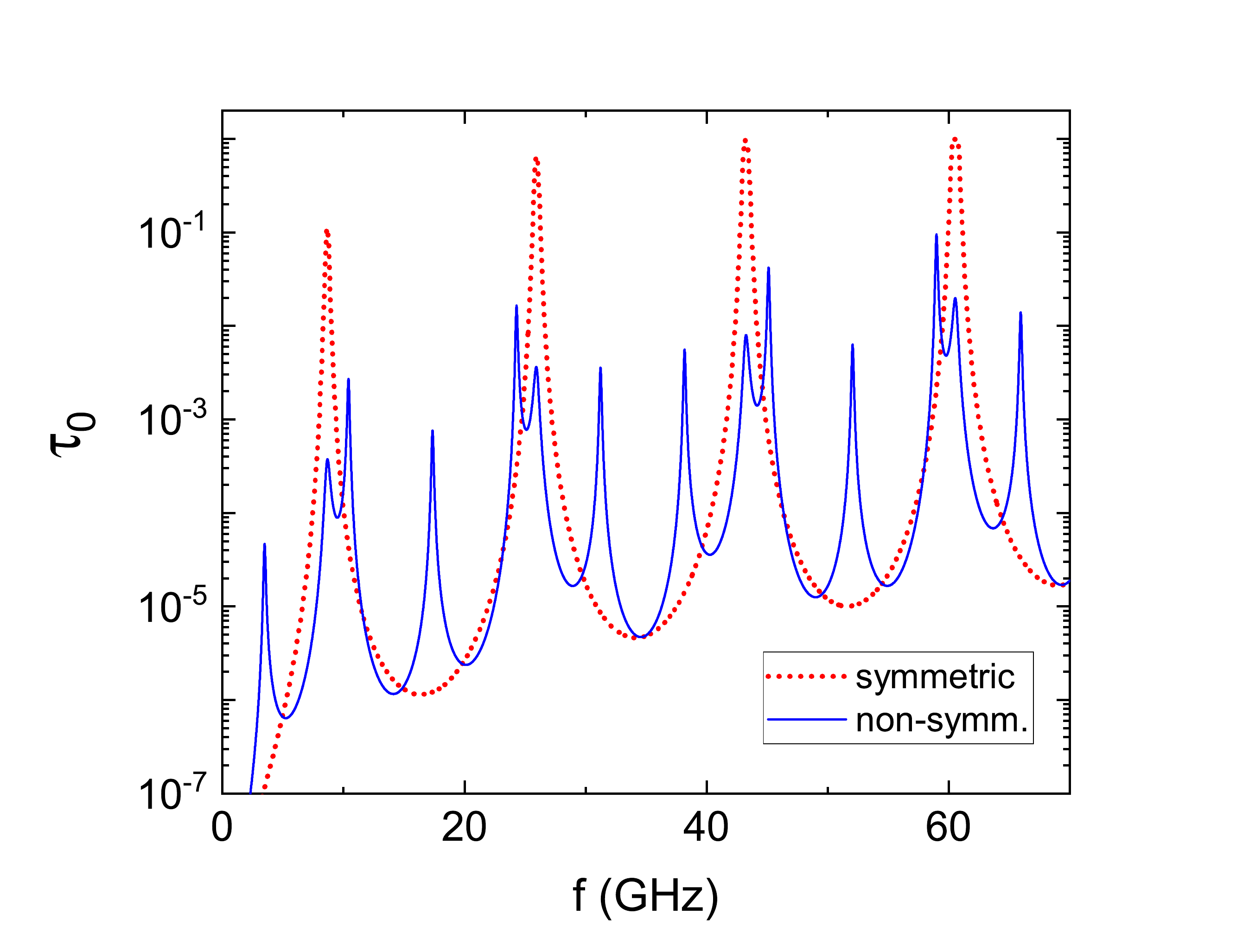}
\caption{Transmission probability (\ref{tau}) of a system with two coupled resonators, shown in Fig. \ref{examples}b, at  $I_C=0$ and $C_1=C_2=15$ fF.
Other parameters are listed in the caption of Fig. \ref{Fig:tau_high_EJ}.
Red dotted line is for symmetric system with identical resonators, $\omega_1/2\pi=\omega_2/2\pi=8.84$ GHz;
blue line represents asymmetric system, $\omega_1/2\pi=8.84$ GHz and $\omega_2/2\pi=3.5$ GHz.  }
\label{Fig:tau0}
\end{figure}

\begin{figure}
\includegraphics[width=\columnwidth]{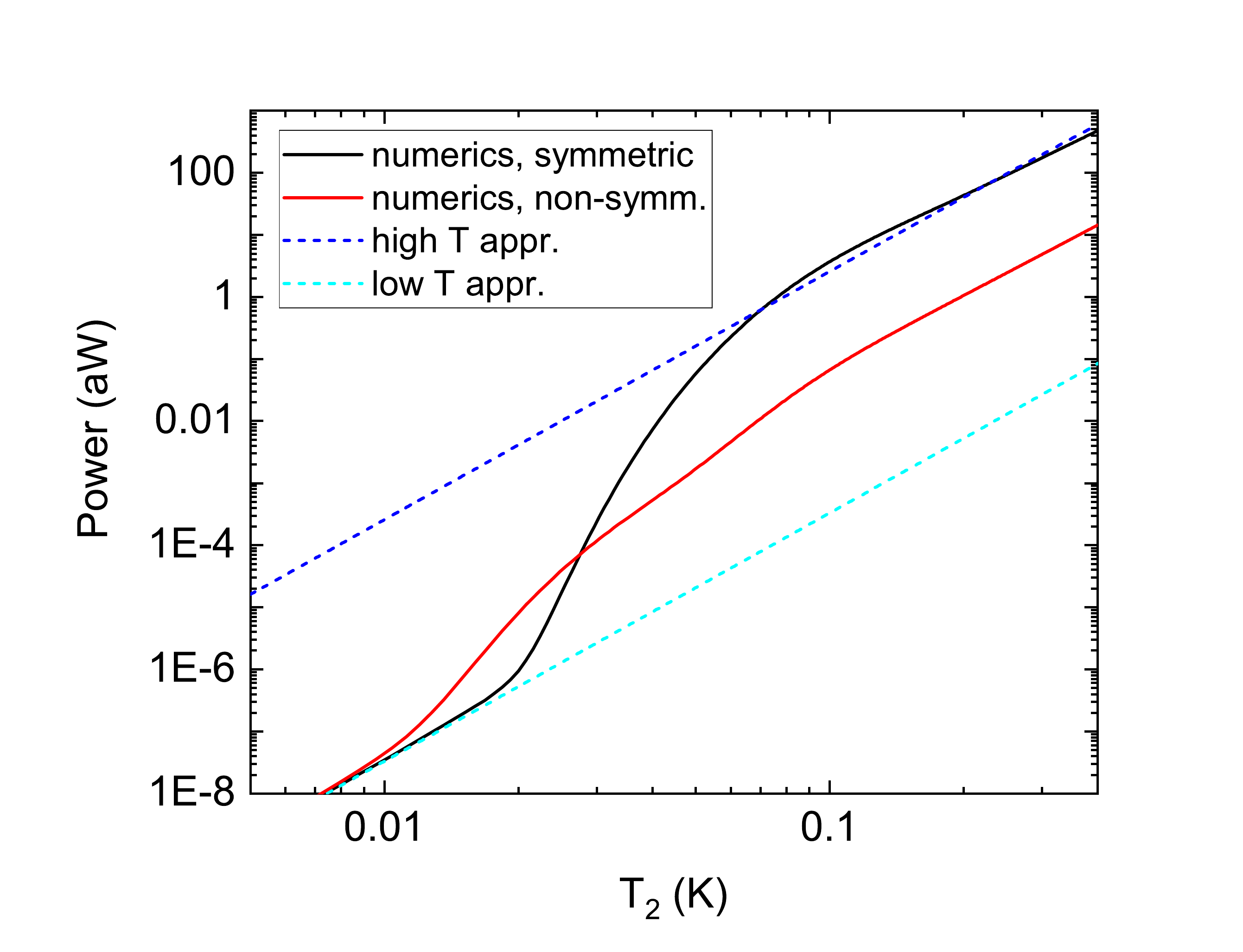}
\caption{Heat flux (\ref{QJ0}) vs the temperature $T_2$ and $T_1=0$ for a system with two coupled resonators and $I_C=0$, $C_1=C_2=15$ fF.
Other parameters are the given in the caption of Fig. \ref{Fig:tau_high_EJ}.
Black solid line represents symmetric system with $\omega_1/2\pi=\omega_2/2\pi=8.84$ GHz;
red solid line is for asymmetric system with $\omega_1/2\pi=8.84$ GHz, $\omega_2/2\pi=3.5$ GHz;
blue dashed line shows high temperature asymptotics (\ref{dotQ_high_T}); 
cyan dashed line is the low temperature expansion (\ref{Q1_cap}). }
\label{no_junction}
\end{figure}

Let us now consider the system with suppressed Josephson current, $I_C=0$.
The transmission probability (\ref{tau}) for such system is plotted in Fig. \ref{Fig:tau0} for 
a symmetric and an asymmetric coupling. In both cases it exhibits sharp peaks at frequencies corresponding to the eigenmodes of the resonators
$\omega_{j,n}=(2n+1)\omega_j$.   
For the case of identical resonators one can derive an accurate high temperature
asymptotics for the integral (\ref{QJ0}) valid at $k_BT_{1,2}\gtrsim \hbar\omega_1/2\pi$.
In this case, one can average the transmission probability (\ref{tau}) over one period $2\omega_1$  
and replace $\tau(\omega)$ in the Eq. (\ref{QJ0}) by its averaged value
\begin{eqnarray}
\langle\tau(\omega)\rangle =  \frac{\frac{1}{2}\left(\frac{Z_{r1}}{R_1}+\frac{R_1}{Z_{r1}}\right)}
{\frac{1}{4}\left(\frac{Z_{r1}}{R_1}+\frac{R_1}{Z_{r1}}\right)^2+\frac{CC_\Sigma}{4C_1^2}\left(\frac{\omega}{\omega_0}+\frac{\omega_0}{\omega}\right)^2},
\end{eqnarray}
where $\omega_0=(Z_{r1}C_1)^{-1}\sqrt{C_\Sigma/C}$. The frequency $\omega_0$ is usually very high so that the condition $k_BT_{1,2}\ll \hbar\omega_0$ is fulfilled.
In this limit one finds 
\begin{eqnarray}
J\approx \Sigma_0\left( \frac{T_2^4}{1+T_2^2/T_0^2} - \frac{T_1^4}{1+T_1^2/T_0^2} \right),
\label{dotQ_high_T}
\end{eqnarray} 
where  the parameter
\begin{eqnarray}
\Sigma_0 = \frac{\pi^3}{15}\frac{Z_{r1}^2C_1^4 k_B^4}{\hbar^3 C_\Sigma^2} \left(\frac{Z_{r1}}{R_1}+\frac{R_1}{Z_{r1}}\right),
\label{Sigma0}
\end{eqnarray}
characterizes the thermal conductance between capacitively coupled resonators, and
\begin{eqnarray}
T_0 = \frac{\sqrt{5}}{\pi\sqrt{2}} \frac{\hbar R_1 C_\Sigma}{k_BC_1^2(Z_{r1}^2+R_1^2)}
\label{T0}
\end{eqnarray}
is the characteristic temperature at which the crossover between two different heat transport regimes occurs.
The result (\ref{dotQ_high_T}) is consistent with the expression (\ref{P1_delta}) in the limit $\omega_J=0$ and $R_1\ll Z_{r1}$.
In Fig. \ref{no_junction} we illustrate the temperature dependence of the heat flux (\ref{QJ0})
for a symmetric system with the two identical resonators and zero Josephson current, $I_C=0$, and
compare it with the approximation (\ref{dotQ_high_T}). We find that the approximation (\ref{dotQ_high_T}) indeed
works quite well at high temperature. The low temperature limit is well described by the Eq. (\ref{Q1_cap}).
For comparison, in the same figure we have also plotted the temperature dependence of the heat flux between two
resonators with different parameters.

\begin{figure}
\includegraphics[width=\columnwidth]{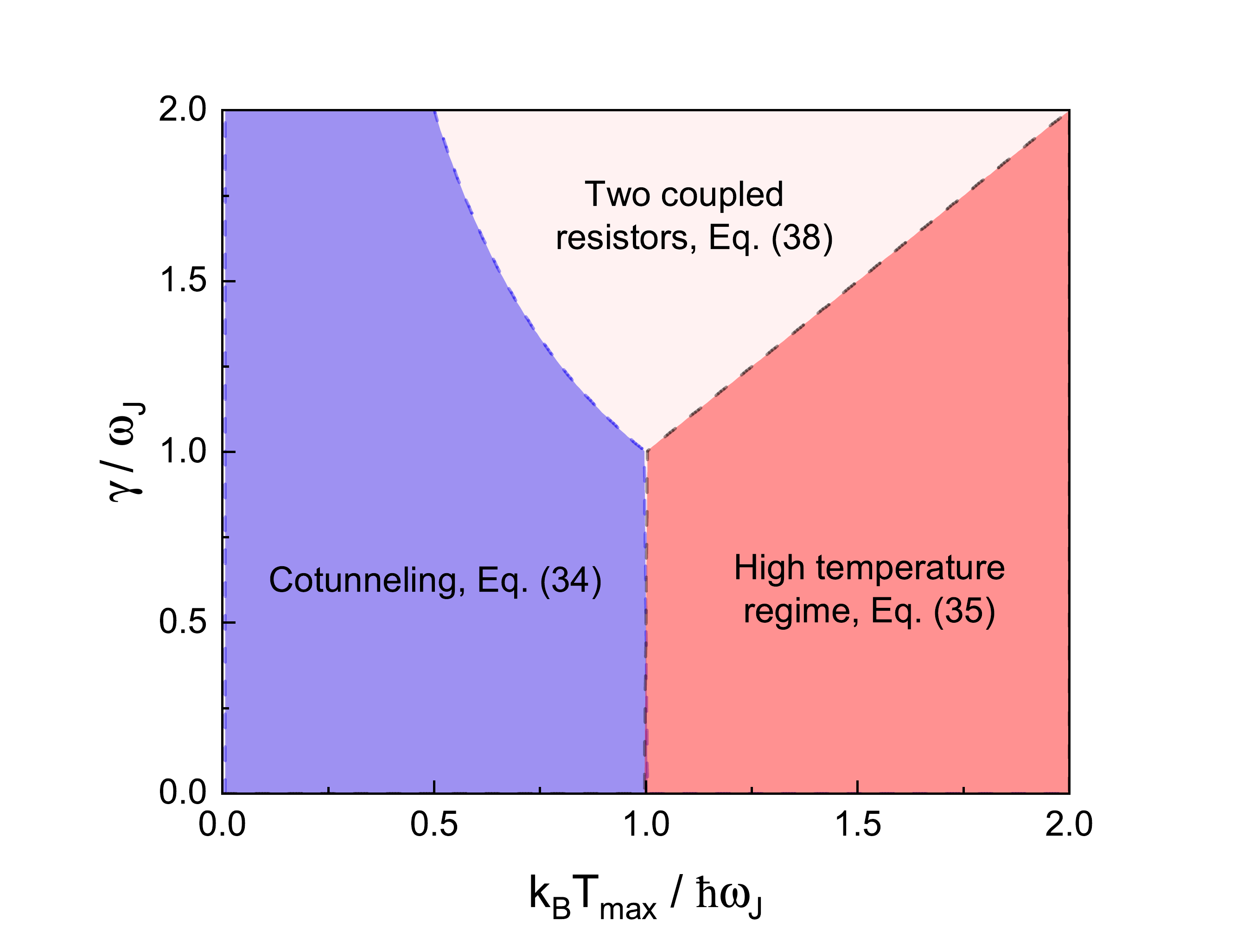}
\caption{Schematic representation of the three heat transport regimes for a linearized Josephson junction coupled to 
two ohmic resistors. Here $T_{\max}=\max\{T_1,T_2\}$.}
\label{diagram}
\end{figure}

Let us now consider the system with ohmic dissipation, $Z_j(\omega)=R_j$. The transmission probability takes the form
\begin{eqnarray}
\tau(\omega)=\frac{4\omega^2}{R_1R_2C^2\left| \omega^2 + i\gamma\omega - \omega_J^2 \right|^2},
\label{tau_Ohmic}
\end{eqnarray} 
where $\gamma=(R_1+R_2)/R_1R_2C$. In the low temperature regime $k_BT_{1,2}\ll \min\{\hbar\omega_J,\hbar\omega_J^2/\gamma\}$ 
one finds the following approximation for the heat flux
\begin{eqnarray}
J = \frac{\pi \alpha_1\alpha_2}{480}\frac{k_B^4(T_2^4-T_1^4)}{\hbar E_J^2},
\label{P1_low_T4}
\end{eqnarray}
where $\alpha_j=h/e^2R_j$ are the dimensionless conductances of the ohmic resistors. 
The result (\ref{P1_low_T4}) may be interpreted as the contribution of co-tunneling, and
it has the same temperature dependence as the co-tunneling contribution to the heat flux through a two level system \cite{Ojanen}.  
At high temperatures $k_BT_{1,2}\gtrsim \max\{ \hbar\omega_J,\hbar\gamma \}$
one finds
\begin{eqnarray}
J = \frac{k_B(T_2-T_1)}{(R_1+R_2)C} - \frac{\hbar}{\pi R_1R_2C^2}\ln\frac{T_2}{T_1}.
\label{P1_high_T}
\end{eqnarray}
Interestingly, this result is independent of $E_J$ even for temperatures $T_{1,2}$ well below the barrier height $2E_J$.
At these temperatures one can roughly approximate the transmission probability (\ref{tau_Ohmic}) by a $\delta$-peak,
\begin{eqnarray}
\tau(\omega) \approx \frac{2\pi}{(R_1+R_2)C} \delta(\omega-\omega_J),
\label{tau_high_T}
\end{eqnarray} 
and the heat flux by the expression
\begin{eqnarray}
J = \frac{\hbar\omega_J[N_2(\omega_J)-N_1(\omega_J)]}{(R_1+R_2)C}.
\label{P1_high_T_2}
\end{eqnarray}
This simple approximation correctly captures the leading term of the high temperature asymptotics (\ref{P1_high_T}), 
but fails to reproduce the sub-leading term as well as the low temperature power law dependence of the heat flux (\ref{P1_low_T4}).
Interestingly,  the $\delta$-peak approximation  (\ref{tau_high_T}) for the transmission probability holds even for an overdamped junction with $\hbar\omega_J\ll \hbar\gamma\ll k_BT_{1,2}$ 
and for a system with $I_C=0$ provided $\hbar\gamma\ll k_BT_{1,2}$. However, in these cases the $\delta$-peak shifts towards zero frequency.
For an overdamped junction with $\gamma \gg \omega_J$ and for $ \hbar\omega_J^2 /\gamma \lesssim k_BT_{1,2}\lesssim \hbar\gamma$
the heat flux (\ref{QJ0}) takes the familiar form 
\begin{eqnarray}
J= \frac{\pi}{12}\frac{4R_1R_2}{(R_1+R_2)^2} \frac{k_B^2(T_2^2-T_1^2)}{\hbar}
\label{P1_low_T}
\end{eqnarray}
describing the heat transfer between two resistors.
In Fig. \ref{diagram} we provide a schematic diagram of the three heat transport regimes discussed above. 

Finally, we consider the limit of sufficiently weak coupling and sufficiently high temperatures
without specifying the frequency dependence of the impedances $Z_j(\omega)$. 
In this regime,  one can replace the transmission (\ref{tau}) by a single $\delta$-peak. 
The corresponding expression reads \cite{SN}
\begin{eqnarray}
\tau(\omega) \approx 
\frac{4\pi E_C}{e^2}
\frac{{\rm Re}\left[ Z_1^{-1}(\omega_J)\right]{\rm Re}\left[Z_2^{-1}(\omega_J)\right]}
{{\rm Re}\left[ Z_1^{-1}(\omega_J) +  Z_2^{-1}(\omega_J) \right]} \delta(\omega-\omega_J).
\nonumber\\ 
\label{delta}
\end{eqnarray}
Afterwards, the heat flux (\ref{QJ0}) acquires a simple form\cite{SN} 
\begin{eqnarray}
J &=& \frac{2\hbar\omega_J E_C}{e^2}
\frac{{\rm Re}\left[Z_1^{-1}(\omega_J)\right]{\rm Re}\left[Z_2^{-1}(\omega_J)\right]}
{{\rm Re}\left[Z_1^{-1}(\omega_J)+Z_2^{-1}(\omega_{J})\right]}
\nonumber\\ &&\times\,
[N_2(\omega_J)-N_1(\omega_J)].
\label{P1_osc_1}
\end{eqnarray}
We test the accuracy of this approximation in  Figs. \ref{anharmonicpart} c and d, where we plot
the heat flux through a system with two identical resonators versus the magnetic flux. 
We find that the approximation (\ref{P1_osc_1}) indeed works very well in the weak coupling regime (Fig. \ref{anharmonicpart}c),
but it fails at stronger coupling (Fig. \ref{anharmonicpart}d).

To conclude this section, we note that replacing the nonlinear Josephson junction
by a linear inductor with the impedance (\ref{ZJ0}) results in a very good approximation for the heat flux
provided the coupling between the junction and the environment is sufficiently strong  and the temperature is sufficiently high.

\section{Effects of anharmonicity in the weak-coupling limit}
\label{weak_coupling}

In this section, we consider the effect of Josephson nonlinearity on the heat flux in 
the weak-coupling limit
\begin{eqnarray}
\,{\rm Re}\left[\frac{R_q}{Z_1(\omega)}+\frac{R_q}{Z_2(\omega)}\right] \ll 1.
\label{WC}
\end{eqnarray}
This condition should be valid for all relevant frequencies $\hbar\omega \lesssim k_BT_{1,2}$.
 It ensures that the transition rates between the eigenstates of the junction Hamiltonian (\ref{HJ}) are smaller than the charging energy $E_C$,
which provides the scale of anharmonicity of the junction.
For a setup with two high-quality resonators, we also require sufficiently strong detuning between the Josephson
frequency and the nearest mode of one of the resonators $\omega_{n_0,j}$ 
\begin{eqnarray}
\frac{\hbar g_j^2\omega_{n_0,j}^3}{\omega_J\omega_j^2|\omega_J-\omega_{n_0,j}|} \ll E_C.
\end{eqnarray} 
In this case the shift of the junction frequency caused by the coupling to the resonators is much smaller than $E_C$. 

Under this conditions, it is convenient to work in the basis of the eigenfunctions $\psi_{qn}(\varphi)$ of the junction Hamiltonian, 
which satisfy Schr\"odinger equation
\begin{eqnarray}
\left[ -4E_C\frac{\partial^2}{\partial\varphi^2} + E_J(1-\cos\varphi) \right]\psi_{qn} = \epsilon_n(q)\psi_{qn}. 
\end{eqnarray}
Here $\epsilon_n(q)$ is the $2e$-periodic energy of the $n-$th Bloch band, which depends on the electric charge $q$ transferred through the junction\cite{Zorin,Averin}.
According to the Bloch theorem the wave functions can be expressed in the form $\psi_{qn}(\varphi)=u_{qn}(\varphi)e^{iq\varphi/2e}$, where
$u_{qn}(\varphi)$ are the $2\pi$-periodic functions normalized as $\int_{-\pi}^\pi d\varphi u_{qm}^*(\varphi)u_{qn}(\varphi)=\delta_{mn}$.
In order to find the transition rates between the energy bands, we need to know the matrix elements of the momentum  $\hat p=-i(\partial/\partial\varphi)$ and of the phase
between the functions $u_{qn}(\varphi)$ belonging to different bands, $m\not=n$,
\begin{eqnarray}
p_{mn}(q) &=& \int_{-\pi}^{\pi} d\varphi \, u^*_{qm}(\varphi)\left(-i\frac{\partial}{\partial\varphi}\right) u_{qn}(\varphi),
\label{pmn}
\\
\varphi_{mn}(q) &=& \int_{-\pi}^{\pi} d\varphi \, u^*_{qm}(\varphi)\varphi u_{qn}(\varphi).
\label{phi_mn}
\end{eqnarray}
Equation (\ref{comm}) leads to the relation between them
\begin{eqnarray}
\varphi_{mn}(q) = \frac{-8iE_C}{\epsilon_m(q)-\epsilon_n(q)}p_{mn}(q).
\end{eqnarray}
We will also use the well known result of solid state physics, which states that the operator $\hat p$ 
within one band acts on any function of $q$ by multiplying it with the combination $(e/4E_C) (\partial\epsilon_n(q)/\partial q)$. 
More detailed information about various analytical approximations 
for the Bloch band energies $\epsilon_n(q)$ and the matrix elements $p_{mn},\varphi_{mn}$ can be found in Refs. \cite{Averin,Koch} and
in  Appendix \ref{WKB}.

With the matrix elements at hand, 
we can express the charge-charge correlation function (\ref{SQ}) as a sum of intraband
[$S_Q^{\rm b}(\omega)$]
and interband [$S_Q^{\rm ib}(\omega)$] contributions,  
\begin{eqnarray}
S_Q(\omega)=S_Q^{\rm b}(\omega) + S_Q^{\rm ib}(\omega). 
\end{eqnarray}
The intraband contribution has the form
\begin{eqnarray}
S_Q^{\rm b} = \frac{e^4}{4E_C^2} \sum_n 
\int dt\,e^{i\omega t}\left\langle \frac{\partial\epsilon_n(\hat q(t))}{\partial q}\frac{\partial\epsilon_n(\hat q(0))}{\partial q} \right\rangle,
\label{Sb}
\end{eqnarray}
while the contribution associated with the interband transitions reads
\begin{eqnarray}
S_Q^{\rm ib} &=&   8\pi\hbar e^2\sum_{m\not= n} \int_{-e}^e dq\, w_n(q) 
\nonumber\\ &&\times\,
\delta(\omega-\omega_{mn}(q))| p_{mn}(q) |^2.
\label{Sib}
\end{eqnarray}
Here we have defined the  interband frequency $\omega_{mn}(q)=[\epsilon_m(q)-\epsilon_n(q)]/\hbar$ and   
the occupation probability $w_n(q)$ of the quantum state described by the wave function $\psi_{nq}(\varphi)$. 
These occupation probabilities are normalized as
\begin{eqnarray}
\sum_{n=0}^\infty \int_{-e}^e dq \, w_n(q) = 1.
\end{eqnarray}
The weak coupling approximation has been explicitly used in deriving Eq. (\ref{Sib}), in which the interaction between the
junction and the thermal baths has been ignored. In contrast, at this stage we keep the intraband correlation function (\ref{Sb}) 
in a general form.

Let us assume that the impedances $Z_j(\omega)$  remain finite in the
low-frequency limit $\omega\to 0$.
In this case the distribution function $w_n(q)$ satisfies the kinetic equation \cite{Zorin,Averin}
\begin{eqnarray}
\frac{\partial w_n}{\partial t} &=& \frac{1}{R_S}\frac{\partial}{\partial q}\left( \frac{\partial \epsilon_{n}(q)}{\partial q} w_n \right)
+ \frac{k_BT_J}{R_S}\frac{\partial^2 w_n}{\partial q^2} 
\nonumber\\ &&
+\, \sum_{m(\not= n)}[\Gamma_{nm}(q)w_m - \Gamma_{mn}(q)w_n]
\label{kinetic}
\end{eqnarray}
valid in the lowest nonvanishing order of the perturbation theory in the interaction Hamiltonians $\hat H_{1,2}^{\rm int}$. 
Here we have introduced the shunt resistance $R_S^{-1}=Z_1^{-1}(0)+Z_2^{-1}(0)$ and the effective temperature of the junction
\begin{eqnarray}
T_J = \frac{Z_1^{-1}(0) T_1 + Z_2^{-1}(0) T_2}{Z_1^{-1}(0) + Z_2^{-1}(0)}.
\label{TJ}
\end{eqnarray}
The interband transition rates, appearing in the last term of Eq. (\ref{kinetic}), are given by the sum of partial contributions of the baths 1 and 2, respectively,
\begin{eqnarray}
\Gamma_{mn}(q) = \Gamma_{mn}^{(1)}(q) + \Gamma_{mn}^{(2)}(q),
\end{eqnarray}
and the latter are defined as
\begin{eqnarray}
\Gamma_{mn}^{(j)}(q) &=& |\varphi_{nm}(q)|^2 
\frac{\hbar\omega_{mn}(q)}{2 e^2}
\nonumber\\ && \times\,
\,{\rm Re}\left[\frac{1}{Z_j(\omega_{mn}(q))}\right] N_j(\omega_{mn}(q)).
\label{Gamma_mn}
\end{eqnarray}
Since the matrix elements $\varphi_{mn}$ do not exceed 1, the condition (\ref{WC}) ensures that
$\Gamma^{(j)}_{mn}+\Gamma^{(j)}_{nm}\ll \hbar|\omega_{mn}|$ and guaranties the validity of the kinetic equation 
(\ref{kinetic}).

Ignoring the interaction effects, we derive the following expressions for the symmetric and antisymmetric parts 
of the intraband correlation function (\ref{Sb}):
\begin{eqnarray}
S_Q^{\rm b}(\omega)+S_Q^{\rm b}(-\omega) &=& \frac{\pi e^4}{E_C^2} \delta(\omega) \int_{-e}^e dq \left(\frac{\partial \epsilon_n}{\partial q}\right)^2 w_n(q),
\nonumber\\
S_Q^{\rm b}(\omega)- S_Q^{\rm b}(-\omega) &=& \frac{\hbar\omega}{2k_BT_J} [S_Q^{\rm b}(\omega)+S_Q^{\rm b}(-\omega) ].
\label{Sb1}
\end{eqnarray}
The first of these equations follows from the conservation of
the charge $q$ in the absence of interactions. As a result, the symmetrized correlator 
$\langle [\partial\epsilon_n(\hat q(t))/\partial q]\,[\partial\epsilon_n(\hat q(0))/\partial q] + [\partial\epsilon_n(\hat q(0))/\partial q]\,[\partial\epsilon_n(\hat q(t))/\partial q]\rangle$
becomes time independent and its Fourier transform reduces to a $\delta-$function peaked at zero frequency.
However, in order to treat low-frequency asymptotics correctly, one should remember that this peak has small finite width. 
The second equation follows from the fluctuation dissipation theorem, which in equilibrium, i.e., at $T_1=T_2=T$, states that
$S_Q^{\rm b}(\omega)- S_Q^{\rm b}(-\omega) = \tanh({\hbar\omega}/{2k_BT}) [S_Q^{\rm b}(\omega)+S_Q^{\rm b}(-\omega) ]$. Since the symmetrized correlator is proportional
to $\delta(\omega)$, the tangent in front of it can be expanded at small frequencies. Afterwards, one can make a replacement $T\to T_J$ because 
in the low frequency Markovian limit the intensities of the noises generated by the two environments become proportional to the corresponding temperatures and also become additive.

Substituting the correlation function (\ref{Sb1}) in the general formula (\ref{Q3}), 
we obtain the heat flux as a sum of the intraband and interband contributions,
\begin{eqnarray}
J = J^{\rm b} + J^{\rm ib}.
\label{Jbib}
\end{eqnarray}
The intraband contribution has the form
\begin{eqnarray}
J^{\rm b} = \frac{T_2-T_1}{Z_2(0)T_1+Z_1(0)T_2}\sum_{n=0}^\infty\int_{-e}^e dq \left(\frac{\partial\epsilon_n}{\partial q}\right)^2 w_n(q),
\label{P1b_weak}
\end{eqnarray}
while the interband part reads
\begin{eqnarray}
J^{\rm ib} &=&   \sum_{n=0}^{\infty}\sum_{m=n+1}^{\infty} \int_{-e}^e dq\; \hbar\omega_{mn}(q)
\nonumber\\ && \times\,
[\Gamma^{(1)}_{nm}(q)w_m(q) - \Gamma^{(1)}_{mn}(q)w_n(q)].
\label{P1ib_weak}
\end{eqnarray}

It is instructive to consider the system in which the frequency dependence of the two bath spectra is the same. 
More precisely, let us assume that
\begin{eqnarray}
{\rm Re}\left[Z_1^{-1}(\omega)\right] = a {\rm Re}\left[Z_2^{-1}(\omega)\right], 
\label{same_bath}
\end{eqnarray}
where $a$ is a frequency independent constant. 
We introduce the heat flux flowing in the direction $1\to 2$, which we will denote as $J_2$. It is given be the expressions (\ref{Jbib},\ref{P1b_weak},\ref{P1ib_weak}) 
with the interchanges indexes $1\leftrightarrow 2$.
Energy conservation in the stationary case implies $J+J_2=0$.
This property is ensured by the kinetic equation (\ref{kinetic}).
Using the energy conservation condition we can 
re-write the heat flux in the form
\begin{eqnarray}
J = \frac{J - a J_2}{1+a}.
\end{eqnarray}
After such symmetrization, well known in the theory of quantum dots \cite{MW,Konig}, 
the total heat flux $J$ acquires the Landauer form (\ref{QJ0}) with the transmission probability 
given by the sum of intraband in interband contributions,
\begin{eqnarray}
\tau(\omega) = \tau^{\rm b}(\omega) + \tau^{\rm ib}(\omega).
\end{eqnarray}
These contributions read
\begin{eqnarray}
\tau^{\rm b}(\omega) &=&  \frac{2\sum_{n=0}^\infty\int_{-e}^e dq \left(\frac{\partial\epsilon_n}{\partial q}\right)^2 w_n(q)}{k_B(Z_2(0)T_1+Z_1(0)T_2)}\delta(\omega),
\label{tau_b}
\\
\tau^{\rm ib}(\omega) 
&=&\frac{\pi\hbar\omega}{e^2}\frac{{\rm Re}[Z_1^{-1}(\omega)]\,{\rm Re}[Z_2^{-1}(\omega)]}{{\rm Re}[Z_1^{-1}(\omega)]+{\rm Re}[Z_2^{-1}(\omega)]}
\nonumber\\ && \times\,
\sum_{n=0}^{\infty}\sum_{m=n+1}^{\infty}\int_{-e}^e dq 
[w_n(q)-w_m(q)]
\nonumber\\ && \times\,
|\varphi_{mn}(q)|^2\delta(\omega-\omega_{mn}(q)).
\label{tau_ib}
\end{eqnarray}
Thus, in the weak coupling limit the intraband transitions produce a narrow low frequency peak in the transmission probability, 
while the interband transitions result a series of transmission bands at higher frequencies. Unlike the transmission probability (\ref{tau}) derived in linear approximation, 
the probabilities (\ref{tau_b}) and (\ref{tau_ib}) may depend on temperatures $T_1$ and $T_2$ via the distribution function $w_n(q)$.  
Hence, in general, $J(T_1,T_2)\not=-J(T_2,T_1)$ and heat rectification becomes possible.
We also note that in the limit $E_J=0$, in which the Bloch bands are reduced to parabolas $(q-2en)^2/2C$,
the intraband transmission probability (\ref{tau_b}) takes the simple form (\ref{tau_high_T}) with $\omega_J=0$,
while the interband contribution (\ref{tau_ib}) vanishes.

Another simple case is the system with strongly asymmetric coupling.
Namely, if we assume that ${\rm Re}[Z_1^{-1}(\omega)]\ll {\rm Re}[Z_2^{-1}(\omega)]$, then the stationary solution of the kinetic equation (\ref{kinetic})
can be found exactly, 
\begin{eqnarray}
w_n(q)=\frac{e^{-\epsilon_n(q)/k_BT_2}}{\sum_{n=0}^\infty \int_{-e}^e dq\, e^{-\epsilon_n(q)/k_BT_2} }.
\label{wn}
\end{eqnarray}
In this regime the junction is thermalized with the thermal bath 2.
For a strongly asymmetric system the expressions for transmission probabilities (\ref{tau_b},\ref{tau_ib}) 
are valid  for arbitrary frequency dependence of the impedances $Z_{j}(\omega)$.

Even though the kinetic equation (\ref{kinetic}) and the heat fluxes (\ref{P1b_weak},\ref{P1ib_weak}) have relatively simple form, the solution
of the problem for an arbitrary ratio $E_J/E_C$ is a challenging task, which requires numerical simulations. In the next two subsections,
we consider the limits $E_J\gg E_C$ and $E_J\ll E_C$, where further approximations are
possible.

\subsection{Tight-binding limit $E_J \gg E_C$.}

\begin{figure}
\includegraphics[width=0.8\columnwidth]{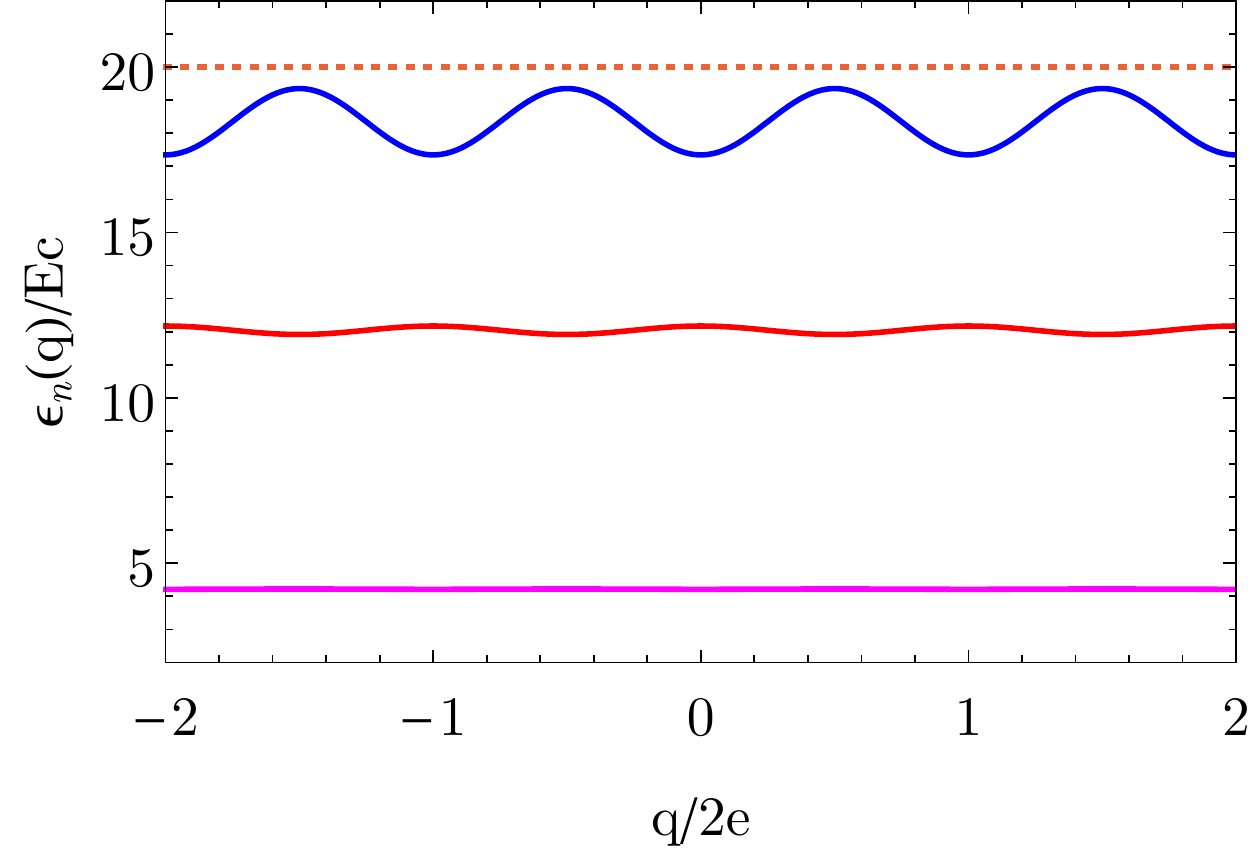}
\caption{The first three Bloch bands (\ref{tb}) of a junction with the following parameters: $E_J/h=1.5$ GHz, 
$E_C/h=150$ MHz. The dotted line indicates the height of the potential barrier $2 E_J$.}
\label{bands}
\end{figure}

In the limit $E_J/E_C\gg 1$,  the lowest Bloch bands acquire cosine dispersion typical
for tight-binding models (see Fig. \ref{bands} for illustration)
\begin{eqnarray}
\epsilon_n(q) = E_n + \delta_n\cos(\pi q/e).
\label{tb}
\end{eqnarray}
Here, $E_n$ is the position of the $n$-th energy level in an isolated potential well and $\delta_n$ is the half-bandwidth
proportional to the hopping amplitude between the wells. Analytical approximations  for both these parameters are summarized
in  Appendix \ref{WKB}. The cosine dispersion (\ref{tb}) is valid for the lowest energy bands with $0 \leq n \leq n_{\max}$, 
where $n_{\max}$, defined by the Eq. (\ref{nmax}), is the number of the highest energy level lying under the barrier $2E_J$.
An example is shown in Fig. \ref{bands}.

\begin{figure}
\includegraphics[width=0.8\columnwidth]{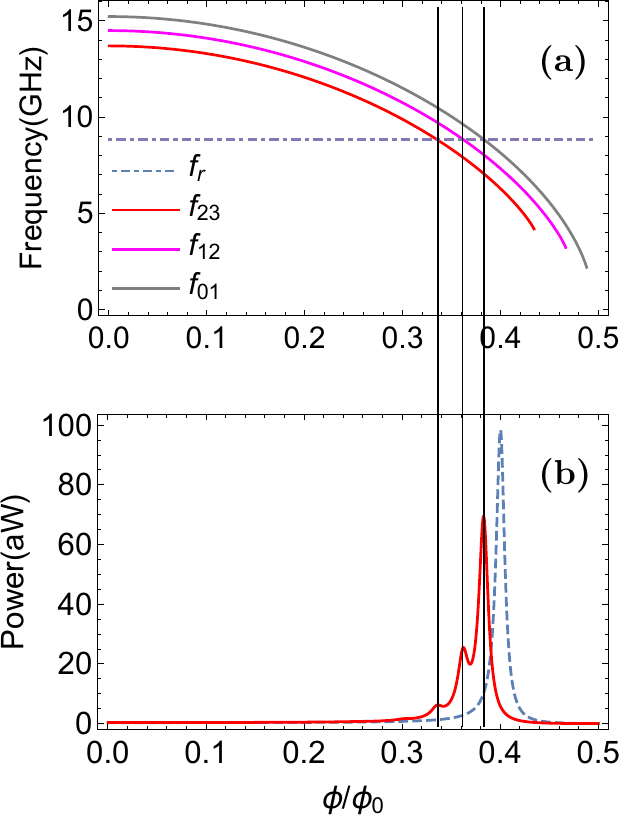}
\caption{Anharmonic effects in the heat flux versus magnetic flux dependence for a system with two identical resonators shown in Fig. \ref{examples}b.
Here, we have used the following parameters: $f_r=\omega_1/(2\pi)=\omega_2/(2\pi)=8.84$ GHz, $E_J/h=48.20 $ GHz, 
$E_C/h=655$ MHz, $T_2=300$ mK, $T_1=150$ mK, $R_1=R_2=2\; \Omega$, $Z_{r1}=Z_{r2}=50\; \Omega$ and $C_1=C_2=1$ fF. 
(a) Frequencies $f_{01}$, $f_{12}$ and $f_{23}$, corresponding to the splitting between the lowest three levels,
as a function of magnetic flux. 
(b) Red line shows the heat flux $J^{\rm ib}$, given by the Eq. (\ref{Q1_weak}), versus the flux $\Phi$. 
Peaks appear when one of the frequencies $f_{j,j+1}$ crosses the frequency of the resonators.
Blue dashed line -- harmonic approximation for the heat flux  (\ref{QJ0},\ref{tau}).}
\label{anharmonicity}
\end{figure}

Next, we assume that $k_BT_{1,2}\lesssim 2E_J$, and
that the coupling between the junction and the environment is not too weak. Namely, we impose the condition
\begin{eqnarray}
\frac{\pi(|\delta_n|+|\delta_{n+1}|)}{(n+1)E_C}\lesssim {\rm Re}\left[\frac{R_q}{Z_1(\omega_J)}+\frac{R_q}{Z_2(\omega_J)}\right] \lesssim 1 
\label{cond1}
\end{eqnarray}
for all bands relevant for the transport of heat, i.e., those with $E_n\lesssim k_BT_{1,2}$.
In this case, the transition rates between the levels (\ref{Gamma_mn}) exceed the half-bandwidth $\delta_n$  and one can put $\delta_n=0$.
Afterwards, the kinetic equation (\ref{kinetic}) acquires a simple form
\begin{eqnarray}
\dot W_n = \sum_{m(\not= n)}[\Gamma_{nm} W_m - \Gamma_{mn} W_n],
\label{master}
\end{eqnarray}
where $W_n=\int_{-e}^e dq \, w_n(q)$ is the total population of the $n-$th band.
The interband  contribution to the heat flux (\ref{P1ib_weak})  is simplified to 
\begin{eqnarray}
J^{\rm ib} =   \sum_{n=0}^{\infty}\sum_{m=n+1}^{\infty}  \hbar\omega_{mn}
[\Gamma^{(1)}_{nm}W_m - \Gamma^{(1)}_{mn}W_n].
\label{Q1_weak}
\end{eqnarray}
The small intraband heat flux, which may dominate at low temperatures, will be analyzed later.
For a system with  baths spectra having the same frequency dependence (\ref{same_bath}), or for a strongly asymmetric system,
the transmission probability  associated with interband transitions (\ref{tau_ib}) acquires the form
\begin{eqnarray}
&& \tau^{\rm ib}(\omega) =\frac{\pi\hbar\omega}{e^2}\frac{{\rm Re}[Z_1^{-1}(\omega)]\,{\rm Re}[Z_2^{-1}(\omega)]}{{\rm Re}[Z_1^{-1}(\omega)]+{\rm Re}[Z_2^{-1}(\omega)]}
\nonumber\\ && \times\,
\sum_{n=0}^{\infty}\sum_{m=n+1}^{\infty} [W_n-W_m]|\varphi_{mn}|^2\delta(\omega-\omega_{mn}).
\label{tau_ib_1}
\end{eqnarray}
Thus every possible transition between the levels results in a separate $\delta$ peak in the transmission probability at frequency $\omega_{mn}$
corresponding to the level spacing.

\begin{figure}
\includegraphics[width=0.8\columnwidth]{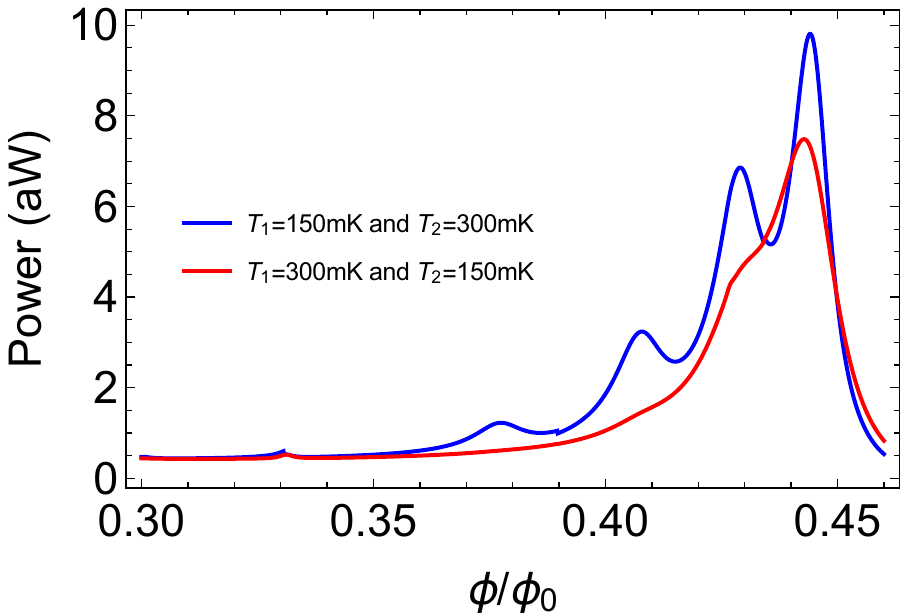}
\caption{Heat rectification in an asymmetric and anharmonic system. 
Here, we have used the following parameters:  $E_J/h=48.20$ GHz, 
 $E_C/h=655$ MHz,  $Z_{r1}=Z_{r2}=50\; \Omega$, $\omega_1/(2\pi)=5.89$ GHz,  $\omega_2/(2\pi)=8.84$ GHz,  $R_1=2\;\Omega$,
 $R_2=20\; \Omega$,  and $C_1= 1$ fF, and $C_2=2$ fF.}
\label{Rectification}
\end{figure}

The energy levels close to the bottom of 
the Josephson potential are not equidistant due to the quartic nonlinearity of the Josephson potential  [see Eq. (\ref{En})],
and the splittings between the lowest neighboring levels vary as $\hbar\omega_{n+1,n}=\hbar\omega_J-(n+1)E_C$.
In Fig. \ref{anharmonicity}a, we plot the dependence of the frequencies $f_{n+1,n}=\omega_{n+1,n}/2\pi$ on the magnetic flux for $n=0,1,2$
for a junction coupled to two identical resonators with the impedances (\ref{Zres}). Every time one of these frequencies crosses the frequency
of the resonators, a peak in the heat flux (\ref{Q1_weak}) appears, as shown in Fig. \ref{anharmonicity}b. 
The height of these peaks is determined by temperatures $T_{1,2}$
via the occupation probabilities $W_n$. At the lowest temperature only one peak, coming from the transitions between the levels 0 and 1, survives.
Since the transmission probability (\ref{tau_ib_1}) depends on temperatures $T_{1,2}$ due to the anharmonicity of the junction, 
one should expect rectification in the system provided the couplings to the two baths are different. In Fig. \ref{Rectification},
we show that this is indeed the case. For the chosen parameters the difference between the heat fluxes $J(T_1,T_2)$ and
$-J(T_2,T_1)$ reaches up to 50\%. 
In order to establish the correspondence with the previous section, we note that
one can find an exact solution of the master equation (\ref{master}) for the junction  with the equidistant
energy levels $E_n=\hbar\omega_J(n+1/2)$ corresponding to the spectrum of a harmonic oscillator, see Appendix \ref{Harmonic_app}. 
For this system the heat flux (\ref{Q1_weak}) takes the form (\ref{P1_osc_1}), while all the peaks
in the transmission probability (\ref{tau_ib_1}) collapse into a single peak and it acquires the form (\ref{delta}).
For comparison, we have plotted the harmonic approximation for the heat flux (\ref{P1_osc_1}) in Fig. (\ref{anharmonicity}).
Clearly, for the chosen parameters it significantly differs from the more accurate result (\ref{Q1_weak}).

For small, but finite, bandwidths $\delta_n$ one can evaluate the integral over $q$ in the general expression for the transmission probability (\ref{tau_ib})
neglecting weak $q$ dependence of the occupation probabilities $w_n$ and matrix elements $\varphi_{mn}$. 
Afterwards, $\tau^{\rm ib}(\omega)$ takes the form (\ref{tau_ib_1}) with the $\delta$ functions replaced by finite width transmission bands
having a double peak shape with square root divergences,
\begin{eqnarray}
\delta(\omega-\omega_{mn}) \to \frac{\hbar}{\pi}\frac{\theta(|\delta_m|+|\delta_n|-|\hbar\omega-E_m+E_n|)}{\sqrt{(|\delta_m|+|\delta_n|)^2-(\hbar\omega-E_m+E_n)^2}}.
\nonumber\\
\end{eqnarray}

Next, we consider the intraband contribution to the heat flux and transmission probability in the tight-binding limit $E_J\gg E_C$. 
For simplicity, we consider only strongly asymmetric systems
with ${\rm Re}\left[Z_1^{-1}(\omega)\right]\ll\, {\rm Re}\left[Z_2^{-1}(\omega)\right]$, for which an analytical solution of the problem is possible.
Taking Boltzmann distribution function (\ref{wn}), in which $\epsilon_n(q)$ have cosine form (\ref{tb}),
we evaluate the intraband heat flux (\ref{P1b_weak}) and arrive at the result
\begin{eqnarray}
J^{\rm b} = \frac{\pi^2}{e^2}\frac{k_B(T_2-T_1)}{Z_1(0)} 
\frac{\sum_{n=0}^{n_{\max}} e^{-\frac{E_n}{k_BT_2}}\delta_n  I_1\left(\frac{\delta_n}{k_BT_2}\right)}
{\sum_{n=0}^{n_{\max}} e^{-\frac{E_n}{k_BT_2}}I_0\left(\frac{\delta_n}{k_BT_2}\right)}.
\label{P1_b}
\end{eqnarray}
Here $I_n(x)$ are modified Bessel  functions.
The intraband heat flux (\ref{P1_b}) may dominate  over the interband one (\ref{Q1_weak}) at low temperatures $k_BT_2\lesssim \hbar\omega_{J}/2\pi$
if $Z_1^{-1}(0)$ is sufficiently large. 
At these temperatures, only the lowest energy band is populated, and we can approximate $J^{\rm b}$ as
\begin{eqnarray}
J^{\rm b} = \frac{\pi^2\delta_0}{e^2Z_1(0)}\frac{I_1\left({\delta_0}/{k_BT_2}\right)}{I_0\left({\delta_0}/{k_BT_2}\right)}k_B(T_2-T_1).
\label{Pb_low}
\end{eqnarray}

One can also derive relatively simple analytical expressions for the heat flux in the temperature interval
$\hbar\omega_{J}/2\pi  \lesssim k_BT_2 \lesssim 2E_J$. 
In this case, the interband contribution can be well approximated by harmonic result (\ref{P1_osc_1}), 
which can be further simplified to the form
\begin{eqnarray}
J^{\rm ib} = \frac{2 E_C}{e^2} {\rm Re}\left[\frac{1}{Z_1(\omega_J)}\right]k_B(T_2-T_1)
\label{P1_osc_app}
\end{eqnarray}
resembling the result (\ref{P1_high_T}) derived for the case of Ohmic dissipation.
The intraband contribution (\ref{P1_b}) in this temperature interval takes the form
\begin{eqnarray}
J^{\rm b} 
\approx \frac{R_q}{4\pi\, Z_1(0) }\frac{T_2-T_1}{\hbar k_B T_2^2}
\frac{\hbar^3\omega_J^3\, e^{-1-{2E_J}/{k_BT_2}}}{\ln\big[16\pi\sqrt{{8E_J}/{E_C}}\big]}.
\label{P1_taps}
\end{eqnarray}
The heat flux (\ref{P1_taps}) is a contribution of thermally activated phase slips, i.e. the jumps between the neighboring
potential wells of the Josephson potential, which become relevant above the quantum to classical crossover temperature of the junction $T^*=\hbar\omega_J/2\pi$. 
The phase slip heat flux (\ref{P1_taps}) 
may dominate over the interband contribution (\ref{P1_osc_app}) if ${\rm Re}[Z^{-1}(0)]\gg {\rm Re}[Z^{-1}(\omega_J)]$.

\subsection{Weak Josephson coupling, $E_J\ll E_C$}

\begin{figure}
\includegraphics[width=0.8\columnwidth]{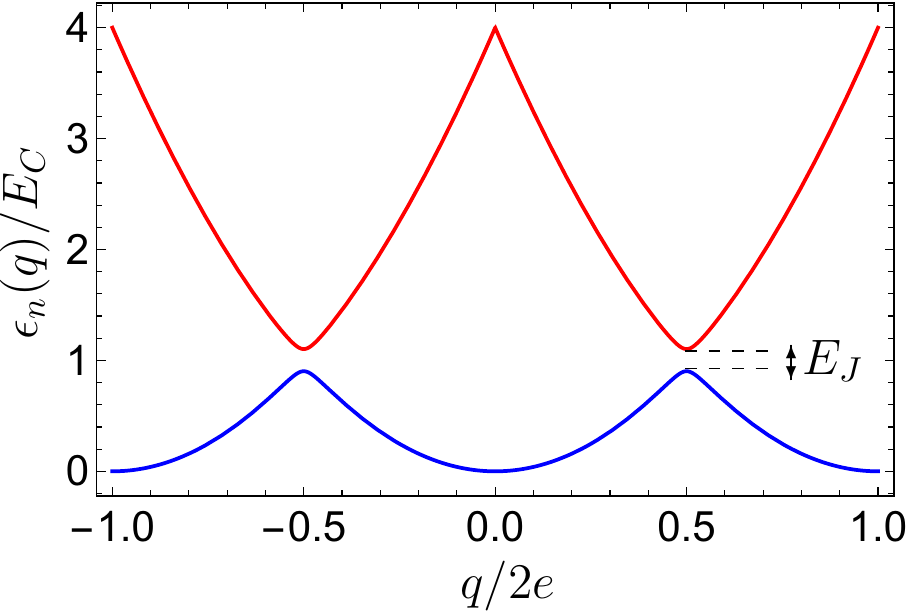}
\caption{The two lowest Bloch bands (\ref{lowestbands}) for the junction parameters $E_J/h=10$ MHz and 
$E_C/h=50$ MHz. 
The energy gap between the two bands equals to $E_J$.}
\label{lowerbands}
\end{figure}

For $E_J\ll E_C$ the energies of the two lowest Bloch bands with $n=0$ and 1 are well approximated by \cite{GR,Jared}
\begin{eqnarray}
\frac{\epsilon_n(q)}{E_C} = \left(\frac{2}{\pi}\arcsin\left[\left(1-\frac{\pi^2}{128}\frac{E_J^2}{E_C^2}\right)\left|\sin\frac{\pi q}{2e}\right|\right]-2n\right)^2.
\nonumber\\
\label{lowestbands}
\end{eqnarray}
These two lowest Bloch bands are illustrated in Fig. \ref{lowerbands}.

Higher lying Bloch bands can be replaced by parabolas $4E_C(n-q/2e)^2$ and small energy gaps between them can be ignored.
Furthermore, the matrix elements of the phase operator connecting the two bands read \cite{Averin} 
\begin{eqnarray}
|\varphi_{01}(q)|^2 = \frac{16E_J^2E_C^2}{\hbar^4\omega_{01}^4(q)}.
\end{eqnarray}

For simplicity, here we only consider a system with strongly asymmetric coupling, for which the distribution function has 
Boltzmann form (\ref{wn}) with the temperature $T_2$. For $E_J\ll E_C$, the contribution of the interband transitions to the
the transmission probability (\ref{tau_ib}) becomes 
\begin{eqnarray}
&&\tau^{\rm ib}(\omega) = \frac{8}{\sqrt{\pi}}\frac{E_J^2E_C^{3/2}}{\sqrt{k_BT_2}}
\frac{{\rm Re}\left[\frac{R_q}{Z_1(\omega)}\right] \sinh\frac{\hbar\omega}{2k_BT_2}}{\hbar^2\omega^2\sqrt{\hbar^2\omega^2-E_J^2}} 
\nonumber\\ &&\times\,
\theta\left(\omega-\frac{E_J}{\hbar}\right)\exp\left[-\frac{E_J+E_C}{k_BT_2} - \frac{\hbar^2\omega^2-E_J^2}{16E_C k_BT_2}\right],
\label{tau_small}
\end{eqnarray}
where $\theta(x)$ is the Heaviside step function.
The transmission probability (\ref{tau_small}) exhibits a gap $E_J/\hbar$ at low frequencies and  gets exponentially suppressed for $k_BT_2\lesssim E_C$.

For temperatures exceeding the gap, $k_BT_{1,2}\gg E_J$, one can derive rather simple analytical expression for the heat flux.
The intraband contribution (\ref{P1b_weak}) takes the form
\begin{eqnarray}
J^{\rm b} = \frac{R_qE_Ck_B(T_2-T_1)}{\pi\hbar Z_1(0)}\left( 1-\frac{\pi-2}{\sqrt{\pi}} \frac{E_J e^{-E_C/k_BT_2}}{k_BT_2}  \right). 
\label{P1b_EJ_ll_EC}
\end{eqnarray}
The leading term in this expression is similar to the heat flux through a capacitor  (\ref{P1_high_T}), while the correction $\propto E_J$
comes from the opening of the energy gap between the two lowest Bloch bands. The interband contribution to the heat flux is small in this limit,
\begin{eqnarray}
J^{\rm ib} &\approx & {\rm Re}\left[\frac{R_q}{Z_1(E_J/\hbar)}\right]\left(\frac{E_C}{k_BT_2}\right)^{3/2} \frac{E_J e^{-E_C/k_BT_2}}{\sqrt{\pi}\hbar} 
\nonumber\\ && \times\,
k_B(T_2-T_1).
\end{eqnarray}

\section{Effects of anharmonicity in the strong-coupling limit}
\label{strong_coupling}

In this section, we consider the strong-coupling limit (\ref{SC}) in more detail
and derive corrections to the results of Sec. \ref{Harmonic} caused by anharmonicity of the junction.
As in the previous section, we will separately consider the regimes $E_J\gg E_C$ and $E_J\ll E_C$.
 
\subsection{Tight-binding limit $E_J\gtrsim E_C$}

In this section, we consider the tight-binding regime $E_J\gtrsim E_C$ and assume that the temperature is low, $k_BT_{1,2}\ll \hbar\omega_J$.
We derive the correction to the Landauer formula (\ref{QJ0}) originating from the finite 
bandwidth of the lowest Bloch band. For this purpose, we perform the expansion in the half-bandwidth $\delta_0$ keeping only the lowest 
nonvanishing term $\propto\delta_0^2$. This correction may also be interpreted as a contribution of 
phase slips, or the jumps between the neighboring potential wells. 
In the weak-coupling limit and for $E_J\gg E_C$ the Landauer heat flux (\ref{QJ0}) crosses over to the interband contribution (\ref{Q1_weak}),
while the phase slip correction $\propto\delta_0^2$ --- to the high-temperature expansion of the intraband contribution (\ref{Pb_low}). 

The phase slip correction is important only at low temperatures $k_BT_{1,2}\ll\hbar\omega_J$ unless the Landauer contribution (\ref{QJ0})
is deliberately made small by specific choice of the circuit impedances $Z_{1,2}(\omega)$.
At these temperatures only the lowest Bloch band is populated and  the intraband charge correlation function (\ref{Sb}) takes the form
\begin{eqnarray}
S^{\rm b}_Q &=& \frac{\pi^2e^2\delta_0^2}{4E_C^2} 
\int dt\, e^{i\omega t}\,\left\langle \sin\frac{\pi\hat q(t)}{e}\sin\frac{\pi\hat q(0)}{e} \right\rangle
\nonumber\\
&=& \frac{\pi^3 e^2\delta_0^2}{4E_C^2}  {\cal P}(\omega).
\label{SQ_b}
\end{eqnarray}
Here,  we have introduced the function
\begin{eqnarray}
{\cal P}(\omega)=\int \frac{dt}{2\pi}e^{i\omega t}\left\langle e^{i\pi\hat q(t)/e} e^{-i\pi\hat q(0)/e} \right\rangle,
\label{cal_P}
\end{eqnarray}
similar to the one appearing in the theory of the Coulomb blockade in a Josephson junction embedded in electromagnetic environment \cite{ANO}.
The function ${\cal P}(\omega)$ has a physical meaning of the probability of photon absorption in either
of the two thermal baths during a phase slip event.
This function should be evaluated at $\delta_0=0$ and, since the temperature is low, one can also ignore interband transitions.
Afterwards, applying the standard methods \cite{IN} and adapting them to the environment consisting of the two baths with different temperatures, 
we obtain ${\cal P}(\omega)$ in the form of the convolution of the two probabilities ${\cal P}_{1,2}(\omega)$
describing photon absorption in the baths 1 and 2,
\begin{eqnarray}
{\cal P}(\omega) = \int d\omega' {\cal P}_1(\omega-\omega'){\cal P}_2(\omega').
\label{convolution_1}
\end{eqnarray}
These probabilities  are given by the integrals
\begin{eqnarray}
{\cal P}_j(\omega)=\int\frac{dt}{2\pi}e^{i\omega t} e^{-{\cal F}_j(t)-i{\cal K}_j(t)},
\label{Pj_q}
\end{eqnarray}
in which the functions ${\cal F}_j(t)$ and ${\cal K}_j(t)$ read
\begin{eqnarray}
{\cal F}_j(t) &=& \frac{\pi\hbar}{e^2}\int_0^\infty d\omega\,{\rm Re}\left[\frac{\omega\coth\frac{\hbar\omega}{2k_BT_j}}{Z_j(\omega)}\right]
\frac{1-\cos\omega t}{\omega^2},
\label{calF}\\
{\cal K}_j(t) &=& \frac{\pi\hbar}{e^2}\int_0^\infty d\omega\,{\rm Re}\left[\frac{1}{Z_j(\omega)}\right]\frac{\sin\omega t}{\omega}.
\label{calK}
\end{eqnarray}
For details of the derivation, we refer the reader to the Appendix \ref{charge}.
The substitution of the correlation function (\ref{SQ_b}) in the general expression (\ref{Q3}) for the heat flux 
results in the phase slip correction (see Appendix \ref{charge} for details)
\begin{eqnarray}
J^{\rm b} = \frac{\pi \delta_0^2}{\hbar} \int d\omega\, \omega\,{\cal P}_1(\omega)\,{\cal P}_2(-\omega).
\label{Pb_final}
\end{eqnarray} 
This expression has a clear physical meaning: the correction to the heat flux is the net contribution of elementary events, in which a photon is
absorbed by the junction from the bath 2 and then re-emitted into the bath 1.
The same expression has been recently derived for a two level system in Ref. \cite{Aurell}.

Interestingly, the correction (\ref{Pb_final}) can be transformed to the Landauer form (\ref{QJ0}) with the aid
of the detailed balance relations
\begin{eqnarray}
{\cal P}_j(\omega)=e^{\hbar\omega/k_BT_j}{\cal P}_j(-\omega).
\label{DB}
\end{eqnarray}
The corresponding contribution to the transmission probability has the form
\begin{eqnarray}
\tau^{\rm b}(\omega) =
\frac{2\pi^2\delta_0^2}{\hbar^2}[{\cal P}_1(\omega)-{\cal P}_1(-\omega)]
[{\cal P}_2(\omega)-{\cal P}_2(-\omega)].
\nonumber\\
\label{tau_bP}
\end{eqnarray}
This expression provides the generalization of the weak-coupling expression for the intraband transmission probability 
(\ref{tau_b}). Indeed, the low frequency $\delta$-peak now acquires a finite width, which  depends
on the bath spectra and may also depend on temperatures $T_{1,2}$.   
The validity condition of the expressions (\ref{Pb_final}) and (\ref{tau_bP}) is $\delta_0\ll k_BT_{1,2}\ll \hbar\omega_J$. 
However, for sufficiently strong coupling, one can use them even at $T=0$.

The computation of the functions ${\cal P}_j(\omega)$ for arbitrary impedances $Z_{1,2}(\omega)$ is 
complicated. However, in the weak-coupling limit (\ref{WC}), one can obtain relatively simple expressions by expanding 
the integrals in  Eqs. (\ref{Pj_q}) to the first order in small functions ${\cal F}_j(t)$ and ${\cal K}_j(t)$. In this approximation, one finds
\begin{eqnarray}
{\cal P}_j(\omega)-{\cal P}_j(-\omega) =   \frac{\pi\hbar}{e^2} \,{\rm Re}\left[\frac{1}{Z_j(\omega)}\right]\frac{1}{\omega}.
\end{eqnarray}
Hence, the correction to the transmission probability (\ref{tau_bP}) simplifies to
\begin{eqnarray}
\tau^{\rm b}(\omega) =
\frac{2\pi^2\delta_0^2}{e^4}
\frac{{\rm Re}\left[Z_1^{-1}(\omega)\right] {\rm Re}\left[Z_2^{-1}(\omega)\right]}{\omega^2}.
\label{tau_PS}
\end{eqnarray}

Let us now consider the two examples, which we have introduced before. First, we consider resonant environments with the impedances (\ref{Zres}).
For high quality-factor resonators with $R_j\ll Z_{rj}$, one can approximate the transmission probability (\ref{tau_PS}) at $\omega\ll \omega_{1,2},\omega_J$ as
\begin{eqnarray}
\tau(\omega) \approx \frac{\hbar^4 R_1R_2C_1^2C_2^2\omega^6}{4e^4E_J^2} + \frac{2\pi^2\delta_0^2}{e^4} R_1R_2C_1^2C_2^2 \omega^2,
\end{eqnarray}
where the first term comes form the transmission probability (\ref{tau}) and the second  from the phase slip correction (\ref{tau_bP}).
The total heat flux takes the form
\begin{eqnarray}
J &=& \frac{\pi^7}{15} \frac{R_1R_2C_1^2C_2^2}{\hbar^3 e^4 E_J^2}\left[(k_BT_2)^8-(k_BT_1)^8\right]
\nonumber\\
&+& \frac{\pi^5}{15} \frac{\delta_0^2R_1R_2C_1^2C_2^2}{\hbar^3 e^4}  \left[(k_BT_2)^4-(k_BT_1)^4\right].
\label{Q1_ind}
\end{eqnarray}
The first term comes from the harmonic approximation  (\ref{QJ0},\ref{tau}) and  dominates at relatively high temperatures $k_BT_{1,2}\gtrsim \sqrt{\delta_0E_J/\pi}$, where
the junction acts as an inductor. In contrast, at low temperatures $k_BT_{1,2}\lesssim \sqrt{\delta_0E_J/\pi}$,
the phase slip contribution (\ref{Pb_final}) dominates the heat transport, and the junction more resembles a capacitor.

Our second example is the case of Ohmic environments with $Z_j(\omega)=R_j$. 
At low frequencies and temperatures $\hbar\omega, k_BT_{1,2}\ll \hbar\omega_J$, the functions (\ref{Pj_q}) acquire the form \cite{IG}
\begin{eqnarray}
{\cal P}_j(\omega) &=& \frac{\hbar e^{\hbar\omega/2k_BT_j}}{4\pi^2 k_BT_j\Gamma(\alpha_j/2)}
\left( \frac{2\pi e^{-\gamma} k_BT_j}{\hbar\omega_J} \right)^{\frac{\alpha_j}{2}}
\nonumber\\ &&\times\,
\left| \Gamma\left( \frac{\alpha_j}{4} + i\frac{\hbar\omega}{2\pi k_BT_j} \right)\right|^2.
\label{calPOhmic}
\end{eqnarray}
Here we assumed that despite strong coupling to the environments the junction remained underdamped, which is possible if
$
\alpha_1+\alpha_2\lesssim \pi\sqrt{{8E_J}/{E_C}}.
$
In this case, one obtains the transmission probability in the form
\begin{eqnarray}
\tau(\omega)=\tau^{\rm b}(\omega) + \tau^{\rm ib}(\omega),
\end{eqnarray}
where $\tau^{\rm b}(\omega)$ is given by  Eq. (\ref{tau_bP}) and 
the interband contribution results form the low frequency  expansion of the transmission (\ref{tau}),
\begin{eqnarray}
\tau^{\rm ib}(\omega) = \frac{4\omega^2}{R_1R_2C^2\omega_J^4}.
\end{eqnarray}
One can now estimate the low temperature heat conductance of the system
\begin{eqnarray}
\kappa =\frac{\partial J}{\partial T_2}\bigg|_{T_1=T_2=T} =\int_0^\infty \frac{d\omega}{2\pi} \frac{\hbar^2\omega^2 \tau(\omega)}{4k_BT^2\sinh^2\frac{\hbar\omega}{2k_BT}}.
\label{kappa0}
\end{eqnarray}
It reads
\begin{eqnarray}
\kappa = \kappa^{\rm ib} + \kappa^{\rm b},
\end{eqnarray}
where the interband part follows from  Eq. (\ref{P1_low_T4})
\begin{eqnarray}
\kappa^{\rm ib} = \frac{\pi \alpha_1\alpha_2}{120} \frac{k_B^4 T^3}{\hbar E_J^2},
\label{kappa_ib}
\end{eqnarray}
and the intraband one is given by
\begin{eqnarray}
\kappa^{\rm b} =\frac{\pi^2 e^{-\gamma}}{2}\frac{\alpha_1\alpha_2}{\alpha_1+\alpha_2}
\frac{k_B\delta_0^2}{\hbar^2\omega_J}\left( \frac{2\pi e^{-\gamma} k_BT }{\hbar\omega_J} \right)^{\frac{\alpha_1+\alpha_2}{2}-1}.
\label{kappa_Ohmic}
\end{eqnarray}
The latter formula is valid for $\alpha_{1,2}\lesssim 4$. One can verify that for $\delta_0\ll k_BT\ll\hbar\omega_J$ and $\alpha_1\ll \alpha_2\ll 1$ 
 Eq. (\ref{kappa_Ohmic}) matches the thermal conductance  in the weak coupling limit (\ref{Pb_low}). It also agrees with the results
of Refs.\cite{Saito,Saito2}, where the heat transport through a two level system has been studied, if one identifies the parameter $\alpha$,
introduced there, with the combination $(\alpha_1+\alpha_2)/4$.

\subsection{Heat transport at $E_J\ll k_BT_{1,2}$}

In this section, we assume that $E_J$ is smaller than the bath temperatures $T_{1,2}$. 
Performing the expansion in the small parameter $E_J/k_BT$, 
we obtain the phase-phase correlation function in the form (see Appendix \ref{Sphi} for details)
\begin{eqnarray}
S_\varphi(\omega) 
= \frac{8e^2}{\hbar\omega}\frac{{\rm Re}\left[ \frac{1+N_1(\omega)}{Z_1(\omega)} + \frac{1+N_2(\omega)}{Z_2(\omega)} \right] 
+ \frac{\pi I_C^2}{2\hbar\omega}P(\omega)}
{\left| -i\omega C + \frac{1}{Z_1(\omega)}+\frac{1}{Z_2(\omega)} + \frac{1}{Z_J(\omega)} \right|^2}.
\nonumber\\
\label{phi_phi}
\end{eqnarray}
Here $Z_J(\omega)$ is the effective impedance of the Josephson junction in presence of strong phase fluctuations \cite{Olli}
\begin{eqnarray}
\frac{1}{Z_J(\omega)} = \frac{I_C^2}{\hbar}\int_0^\infty dt \frac{1-e^{i\omega t}}{-i\omega} e^{-F(t)}\sin[K(t)],
\label{ZJ1}
\end{eqnarray}
and 
\begin{eqnarray}
P(\omega) &=& \int \frac{dt}{2\pi}\, e^{i\omega t}\,\left\langle e^{i\hat\varphi(t)} e^{-i\hat\varphi(0)}  \right\rangle 
\nonumber\\
&=& \int d\omega\, P_1(\omega-\omega')P_2(\omega')
\label{P}
\end{eqnarray}
is the probability for the junction to emit a photon with the frequency $\omega$ into the environment.
This function resembles the function ${\cal P}(\omega)$ defined in Eq. (\ref{cal_P}), 
but differs from it because the charge $q$ and the phase $\varphi$ fluctuate in a  different way.
The absorption probabilities of the two baths, $P_j(\omega)$, are given by  Eqs. (\ref{Pj_q}) with the functions ${\cal F}_j(t)$, ${\cal K}_j(t)$ replaced 
by similar functions describing phase fluctuations, 
\begin{eqnarray}
F_j(t) &=&  
\frac{4e^2}{\pi\hbar}\int_0^\infty  d\omega\,{\cal R}_j(\omega)\coth\frac{\hbar\omega}{2k_BT_j}\frac{1-\cos\omega t}{\omega},
\label{FKF}\\
K_j(t) &=& 
\frac{4e^2}{\pi\hbar}\int_0^\infty d\omega\, {\cal R}_j(\omega)\frac{\sin\omega t}{\omega}.
\label{FKK}
\end{eqnarray} 
Here the effective spectra of the environments ${\cal R}_j(\omega)$ are
\begin{eqnarray}
{\cal R}_j(\omega) = \frac{{\rm Re}\left[Z_j^{-1}(\omega)\right]}{\left| -i\omega C + Z_1^{-1}(\omega) + Z_2^{-1}(\omega)   \right|^2}.
\end{eqnarray}
The functions $F_j(t)$ and $K_j(t)$ are well known from the theory of environmental Coulomb blockade\cite{IN,Saclay,Falci}.

The substitution of the correlation function (\ref{phi_phi}) in the general expression for the heat flux (\ref{Q30}) gives the result
\begin{eqnarray}
J = J_{1}^{\rm el} + J_{1}^{\rm in},
\label{P_total}
\end{eqnarray} 
where the elastic contribution to the heat flux $J_{1}^{\rm el}$ is given by Landauer formula (\ref{QJ0}) with the transmission probability (\ref{tau}) 
containing the modified junction impedance (\ref{ZJ1}), and $J_{1}^{\rm in}$ describes the contribution coming from the inelastic scattering of photons on the junction,
\begin{eqnarray}
J^{\rm in} = \frac{\pi E_J^2}{\hbar}\int d\omega\, \omega\, P_1(\omega)P_2(-\omega).
\label{Pin}
\end{eqnarray}
The inelastic heat flux (\ref{Pin}) has the same form as the intraband contribution (\ref{Pb_final}), which is the manifestation
of the phase-charge duality well known in the theory of Josephson junctions.
 
The probabilities $P_j(\omega)$ also satisfy detailed balance relations (\ref{DB}),
which allows us to express the total heat flux (\ref{P_total}) in the Landauer form (\ref{QJ0})
with the transmission probability given by the sum of elastic and inelastic contributions,
\begin{eqnarray}
\tau(\omega)=\tau^{\rm el}(\omega) + \tau^{\rm in}(\omega). 
\end{eqnarray}
The elastic part $\tau^{\rm el}(\omega)$ is given by Eq. (\ref{tau}) with $Z_J^{-1}(\omega)$ defined in Eq. (\ref{ZJ1}), and the inelastic one reads
\begin{eqnarray}
\tau^{\rm in}= \frac{2\pi^2 E_J^2}{\hbar^2}[P_1(\omega)-P_1(-\omega)][P_2(\omega)-P_2(-\omega)].
\label{tau_EJ}
\end{eqnarray}

\begin{figure}
\includegraphics[width=\columnwidth]{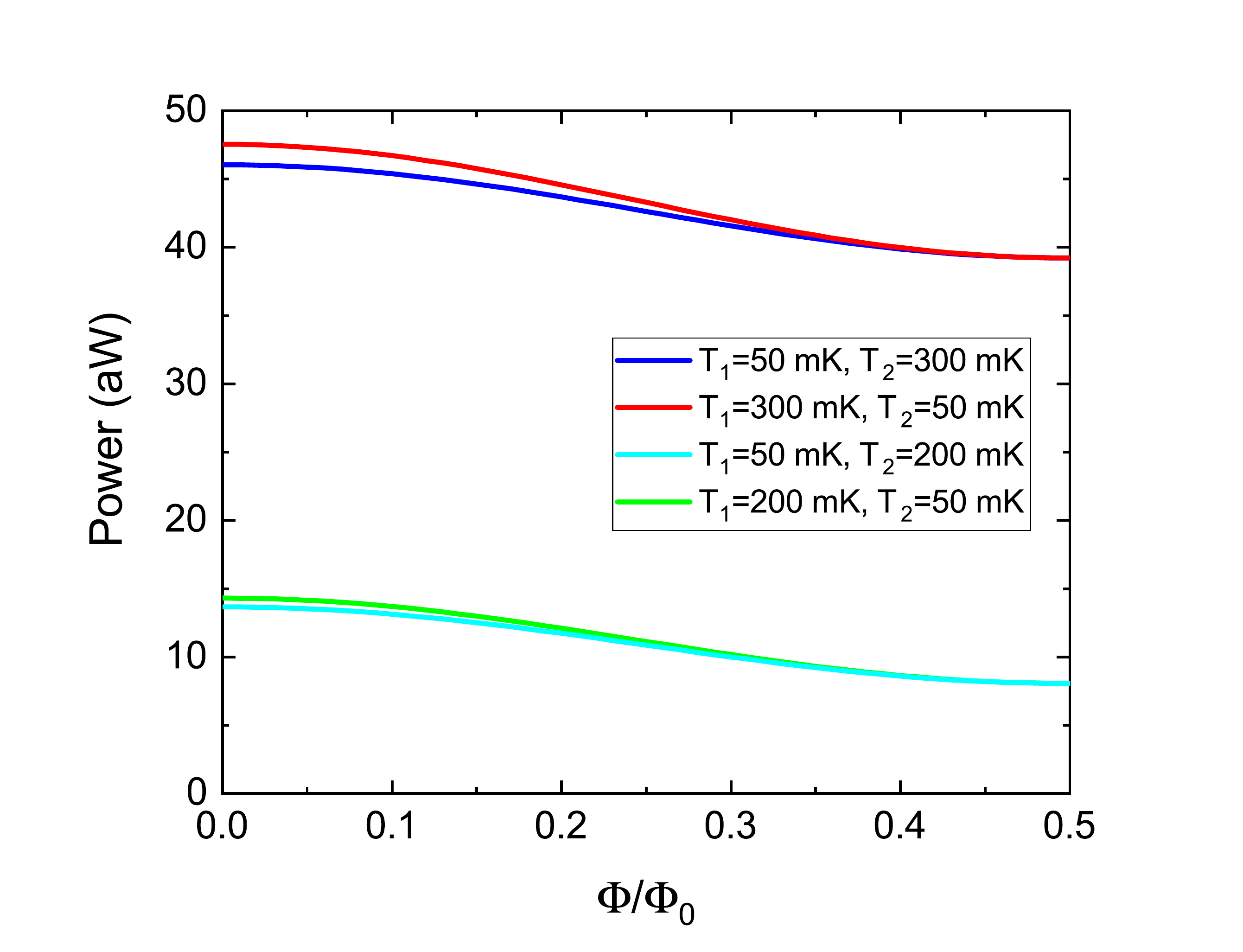}
\caption{Heat flux vs magnetic flux in the regime $E_J\ll k_BT_{1,2}$.
The parameters of the system are: $Z_{r1}=Z_{r2}=50$ $\Omega$, $R_1=1$ $\Omega$, $R_2=30$ $\Omega$, $C_1=5$ fF, $C_2=10$ fF, $C=1$ fF,
$E_C/h = 1.21$ GHz, $E_J=0.7 E_C=0.847$ GHz, and $\omega_1/2\pi = \omega_2/2\pi = 5$ GHz.}
\label{small_EJ}
\end{figure}

As in the previous sections, we consider two examples.
First,  we assume that the junction is coupled to the two resonators with the impedances (\ref{Zres}) and that the coupling is sufficiently weak, $|{\cal R}_j(\omega)|\lesssim R_q$.  
In this case, one can derive the following approximations for the functions $P_j(\omega)$ and the junction impedance (\ref{ZJ1}):
\begin{eqnarray}
&& P_j(\omega)-P_j(-\omega) = \frac{4e^2}{\pi\hbar}\frac{\omega}{\omega^2+\Gamma_{j,T}^2}{\cal R}_j(\omega),
\nonumber\\
&& \frac{1}{Z_J(\omega)} = \frac{2e^2I_C^2}{-i\hbar^2\omega }\left[
\frac{Z_S(i\Gamma_T)}{\Gamma_T} - \frac{Z_S(\omega+i\Gamma_T)}{\Gamma_T-i\omega}\right].
\label{ZJ_app}
\end{eqnarray}
Here, we have introduced the effective total impedance of the environment seen by the junction
\begin{eqnarray}
Z_S=\frac{1}{-i\omega C + Z_1^{-1} + Z_2^{-1}} - \frac{1}{-i\omega C_\Sigma},
\end{eqnarray} 
and used the approximation $F_j(t)=\Gamma_{j,T} |t|$ for the functions (\ref{FKF}). 
The rates of phase diffusion are  
\begin{eqnarray}
\Gamma_{j,T} = \frac{4e^2k_B}{\hbar^2C_\Sigma^2}R_jC_j^2 T_j.
\end{eqnarray}
Equation (\ref{ZJ_app}) contains the total phase diffusion rate
\begin{eqnarray}
\Gamma_T=\Gamma_{1,T}+\Gamma_{2,T} = \frac{4e^2k_B}{\hbar^2}\frac{R_1C_1^2 + R_2C_2^2}{C_\Sigma^2} T_J,
\end{eqnarray}
where the effective temperature of the junction is
\begin{eqnarray}
T_J = \frac{R_1C_1^2T_1 + R_2C_2^2T_2}{R_1C_1^2 + R_2C_2^2}.
\end{eqnarray}
The approximation (\ref{ZJ_app}) is valid if
\begin{eqnarray}
E_J\ll k_BT_J.
\label{cond2}
\end{eqnarray} 

In Fig. \ref{small_EJ}, we show the dependence of the heat flow (\ref{P_total}) on magnetic flux for a certain 
choice of system parameters. For these parameters the effect of the junction impedance (\ref{ZJ1}) on the transmission
probability (\ref{tau}) is insignificant, and
the cosine flux dependence of the heat power predominantly comes from the pre-factor $E_J^2\propto \cos^2(\pi\Phi/\Phi_0)$
in front of the inelastic term (\ref{Pin}). 
We have chosen an asymmetric coupling, that is why the heat flux (\ref{Pin}) exhibits weak rectification.
At low temperatures $k_BT_{1,2}\ll\hbar\omega_{1,2}$, the inelastic contribution to the heat flux (\ref{Pin}) can be estimated as
\begin{eqnarray}
J^{\rm in}\approx \frac{I_C^2 R_1R_2C_1^2C_2^2(T_2-T_1)}{2C_\Sigma^2(R_1C_1^2T_1 + R_2C_2^2T_2)}.
\end{eqnarray}
Figure \ref{small_EJ} also shows that magnetic field independent background heat flux has strong temperature dependence. 
This background comes from the heat transport between two capacitively coupled resonators 
(\ref{Q1_cap},\ref{dotQ_high_T}).

Our second example is again the case of Ohmic dissipation.
In this case, one can derive analytic expressions for the photon absorption probabilities \cite{IG} $P_j(\omega)$  and the effective impedance of the junction (\ref{ZJ1}),
\begin{eqnarray}
P_j(\omega) &=& \frac{\hbar e^{\hbar\omega/2k_BT_j}}{4\pi^2 k_BT_j\Gamma(8\alpha_j/\alpha_\Sigma^2)}
\left( \frac{2\pi^2 e^{-\gamma} k_BT_j}{\alpha_\Sigma E_C} \right)^{\frac{8\alpha_j}{\alpha_\Sigma^2}}
\nonumber\\ &&\times\,
\left| \Gamma\left( \frac{4\alpha_j}{\alpha_\Sigma^2} + i\frac{\hbar\omega}{2\pi k_BT_j} \right)\right|^2,
\label{POhmic}
\\
\frac{1}{Z_J(\omega)} &=& \frac{\pi I_C^2}{2\hbar\omega}\bigg[ P(\omega)-P(-\omega) 
\nonumber\\ &&
-\, i \big(  P(\omega) + P(-\omega) - 2P(0) \big)\tan\frac{4\pi}{\alpha_\Sigma} \bigg],
\label{ZOhmic}
\end{eqnarray}
where $\alpha_\Sigma=\alpha_1+\alpha_2$.
The expression (\ref{POhmic}) is valid provided $\pi k_BT_{1,2}\ll \alpha_\Sigma E_C$ and the approximation for
the impedance (\ref{ZOhmic}) requires the temperatures to be in the interval $E_J \ll k_BT_{1,2}\ll \alpha_\Sigma E_C/\pi$. 
Within this model one can analytically derive the correction to the thermal conductance.
Assuming  that $\alpha_\Sigma\gtrsim 8$, we find
\begin{eqnarray}
\kappa \approx \frac{2\pi}{3} \frac{\alpha_1\alpha_2}{\alpha_\Sigma^2}\frac{k_B^2 T}{\hbar}
- \frac{4\pi \alpha_1\alpha_2}{\alpha_\Sigma^3} \frac{E_J^2}{\hbar T} \left( \frac{2\pi^2 e^{-\gamma} k_BT}{\alpha_\Sigma E_C} \right)^{\frac{8}{\alpha_\Sigma}}.
\nonumber\\
\label{kappa}
\end{eqnarray}
The first term in this expression comes from  the heat current at $E_J=0$ given by  Eq. (\ref{P1_low_T}), and
the second term provides a negative correction to it.  This correction is a combined effect of the inelastic contribution (\ref{Pin}), which gives
positive correction to $\kappa$, and of the junction impedance $Z_J(\omega)$, which suppresses the transmission 
probability together with the thermal conductance.

\section{Summary}
\label{conclusion}

We have studied the photonic heat transport across a Josephson junction coupled to the two linear electric circuits, 
acting as thermal baths, which are
characterized by the impedances $Z_1(\omega),$ $Z_2(\omega)$ and by temperatures $T_1$, $T_2$.
We have shown that linear approximation, in which the nonlinear junction is replaced by an inductor,  
provides a rather good estimate of the heat flux between the thermal baths 
for any relation between the parameters $E_J,E_C, k_BT_{1,2}$, for any coupling strength and for any frequency dependence of
the impedances $Z_{1,2}(\omega)$. This approximation fully accounts for hybridization between the modes
of the junction and of thermal baths.

However, simple harmonic approximation cannot capture subtle effects
such as thermal rectification, for example.
Therefore we have developed more elaborate approximations, which take into account the nonlinear nature of the Josephson junction. 
In the weak-coupling limit, the photon transmission probability of the system is determined by
the Bloch band structure of the junction energy spectrum. 
At low frequencies it is predominantly determined by the transitions between the junction states within one energy band, 
while at high frequencies -- by the transitions between different bands.  

In the limit $E_J\gg E_C$ and at weak coupling  
the photon transmission probability is given by a series of narrow peaks 
with the positions corresponding to the splitting between the energy levels in the Josephson potential well. 
At stronger coupling between the junction and the environment, these peaks overlap forming a single broad peak. 
An additional peak associated with the intraband transitions is formed at low frequencies. 
Its shape is determined by the impedances $Z_{1,2}(\omega)$,  see Eq. (\ref{tau_bP}).
The heat flux at strong coupling is given by the sum of 
harmonic contribution (\ref{QJ0}) and the contribution of phase slips (\ref{Pb_final}), with the latter 
dominating at low temperatures.

In the opposite limit, $E_J\ll E_C$, the Bloch bands become wide and the energy gaps between them almost vanish. 
In this case, the dominating contribution to the heat flux comes from the Landauer formula (\ref{QJ0}), in which one should put $I_C=0$.
In addition, there exists a small correction depending on the Josephson energy $E_J$.  
If $E_J\ll k_BT_{1,2}$ this correction can be split into two parts:  the elastic one, coming from the junction impedance averaged over phase
fluctuations (\ref{ZJ1}), and the inelastic correction (\ref{Pin}), which is associated with the absorption and re-emission of the photons
by the junction. 

Many of our predictions can be experimentally tested. 
Indeed, with reasonable values of the system parameters we have obtained the heat
fluxes in the range 10 - 100 aW (0.03\%-0.3\% of the maximum  value corresponding to a single thermal conductance quantum), see, e.g., Figs. \ref{anharmonicpart} and \ref{anharmonicity}b. 
Such heat fluxes can be reliably detected\cite{Alberto}. This opens up a possibility  
for quantum thermodynamics experiments with tunable Josephson junctions.

\section{Acknowledgement}

We acknowledge useful discussions with Keiji Saito, Erik Aurell and Bayan Karimi.
This work was supported by the Academy of Finland Centre of Excellence
program (project 312057) and the European Union's Horizon 2020 research and innovation programme under the European Research Council (ERC) programme (grant agreement 742559).

\appendix

\section{Heat transport and linearized Langevin equation}
\label{sec-Langevin}

In this appendix, we will demonstrate how one can derive the Landauer formula for the heat current (\ref{QJ0})
by solving the Langevin equations, which exactly describe quantum dynamics of a linear system \cite{Schmid,Dhar}.
Here, we follow Nyquist \cite{Nyquist} and Pascal, Courtois, and Hekking \cite{Hekking}. 

\begin{figure}[!ht]
\includegraphics[width=0.9\columnwidth]{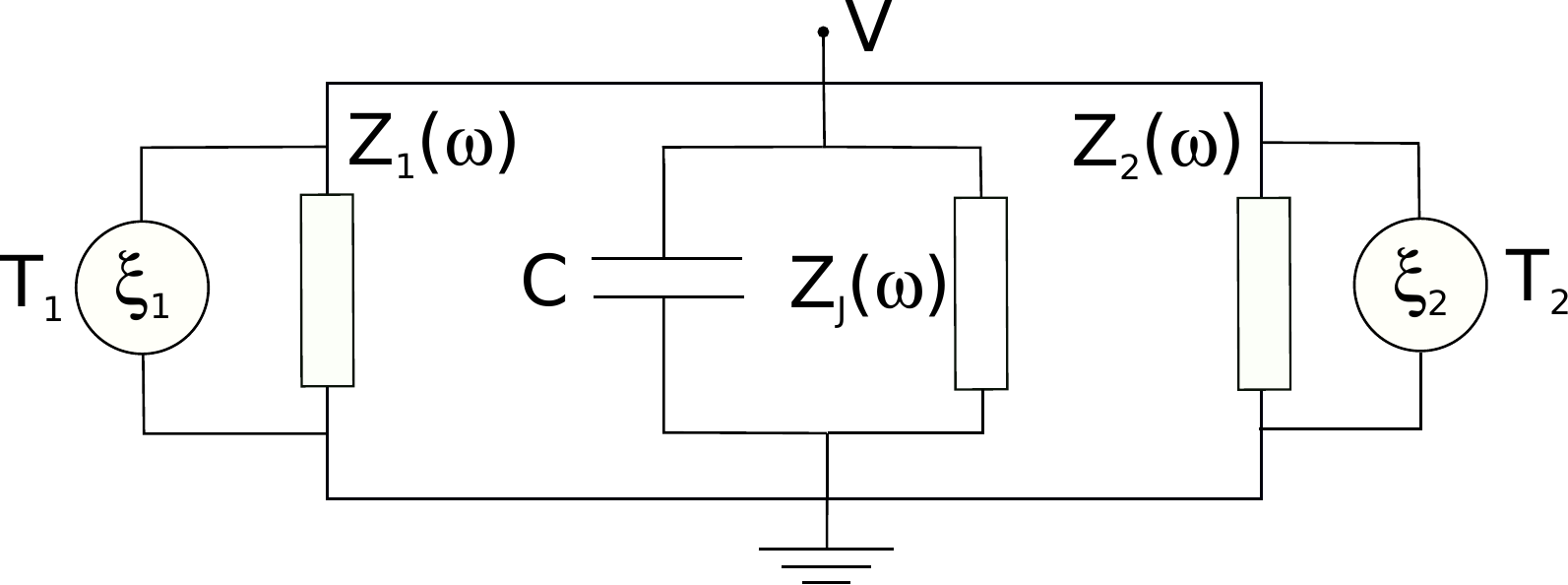}
\caption{Equivalent circuit representation of the system shown in Fig. \ref{schematics} with the SQUID been replaced by
a linear lumped element with the impedance $Z_J(\omega)$. Current sources generating the noise currents $\xi_1(t)$ and 
$\xi_1(t)$, which carry the information about the temperatures $T_1,T_2$.}
\label{Langevin}
\end{figure}

Kirchhoff's equations for the circuit of Fig. \ref{Langevin} read
\begin{eqnarray}
&& I_1(t) = \int_{-\infty}^t dt' Y_1(t-t') V(t') + \xi_1(t),
\label{I1}\\ 
&& I_2(t) = \int_{-\infty}^t dt' Y_2(t-t') V(t') + \xi_2(t),
\label{I2}\\
&& I_J(t) = C\dot V(t) + I_C\sin\varphi,
\label{IJ}
\nonumber\\
&& I_1(t)+I_2(t)+I_J(t) = 0.
\label{conserve}
\end{eqnarray}
Here the admittances of the environments 1 and 2 in the time domain are defined as $Y_j(t)=\int d\omega\, e^{-i\omega t}/2\pi Z_j(\omega)$,
the voltage drop $V$ across the junction is related to the phase by means of the Josephson relation $V=\hbar\dot\varphi/2e$, 
the noises $\xi_1$ and $\xi_2$ are the Gaussian stochastic processes fully characterized by their pair correlators
\begin{eqnarray}
\langle\xi_i(t')\xi_j(t'')\rangle = \int\frac{d\omega}{2\pi} \, \langle |\xi_j|^2_\omega \rangle \,\cos[\omega(t'-t'')]\delta_{ij},
\label{noises}
\end{eqnarray}
\begin{eqnarray}
\langle |\xi_j|^2_\omega \rangle = \,{\rm Re}\left[\frac{1}{Z_j(\omega)}\right]\hbar\omega\coth\frac{\hbar\omega}{2k_BT_j}.
\label{xi_j}
\end{eqnarray}
The currents $I_1(t)$ and $I_2(t)$ flow through the impedances $Z_1(\omega)$ and $Z_2(\omega)$ respectively.
The heat flux in this framework is given by the Joule heating released in the circuit 1 and averaged over the noises $\xi_j$,
\begin{eqnarray}
J = \langle I_1V \rangle_{\xi}.
\label{Q1app}
\end{eqnarray}

Equations (\ref{I1}-\ref{conserve}) can be readily solved by means of
Fourier transformation. We find the Fourier component of the voltage, $V_\omega = \int dt\, e^{i\omega t}\, V(t)$ in the form
\begin{eqnarray}
V_\omega = -\frac{\xi_{1,\omega}+\xi_{2,\omega}}{-i\omega C + \frac{1}{Z_1(\omega)} + \frac{1}{Z_2(\omega)} + \frac{1}{Z_J(\omega)} }.
\label{V}
\end{eqnarray} 
The heat flux (\ref{Q1app}) can now be transformed as
\begin{eqnarray}
J &=& \int\frac{d\omega}{2\pi} \langle I_{1,\omega}V_{-\omega} \rangle_\xi
\nonumber\\
&=& \int\frac{d\omega}{2\pi} \left\langle\left( \frac{V_\omega}{Z_1(\omega)} + \xi_{1,\omega}\right)V_{-\omega}\right\rangle_\xi.
\nonumber
\end{eqnarray}
Combining this expression with  Eq. (\ref{V}), we obtain 
\begin{eqnarray}
J = {\rm Re}\int\frac{d\omega}{2\pi}
\frac{\frac{\langle|\xi_2|^2_\omega\rangle}{Z_1} - \frac{\langle|\xi_1|^2_\omega\rangle}{Z_J} - \frac{\langle|\xi_1|^2_\omega\rangle}{Z_2}}
{\left| -i\omega C + \frac{1}{Z_1} + \frac{1}{Z_2} + \frac{1}{Z_J}  \right|^2}.
\label{QJ}
\end{eqnarray}
Using the expressions (\ref{xi_j}) for the spectral power of the noises, one can show that this expression is equivalent to the Landauer formula (\ref{QJ0})
with the transmission probability having the form (\ref{tau}).

\section{WKB approximation in the limit $E_J\gtrsim E_C$}
\label{WKB}

Let us consider the junction with  $E_J\gtrsim E_C$. It is well known that in this regime
the eigen-energies of the $2\pi$-periodic junction Hamiltonian $\hat H_J$ form Bloch bands with the cosine dispersion (\ref{tb}).
The approximation (\ref{tb}) applies
if $E_n<2E_J$, i.e., for the wave functions localized inside the potential wells and having eigenenergies smaller than the height of the barrier separating neighboring wells.
In this appendix, we summarize approximate analytical expressions for both 
energy levels $E_n$ and half-bandwidth $\delta_n$ of the corresponding Bloch bands, which can be derived by means of the Wentzel-Kramers-Brillouin (WKB) approximation. 

The energies of the quasiclassical levels $E_n^{\rm qcl}$ are determined by the Bohr-Sommerfeld quantization rule,
\begin{eqnarray}
\int_{-\varphi_0(E_n^{\rm qcl})}^{\varphi_0(E_n^{\rm qcl})} d\varphi \frac{\sqrt{E_n^{\rm qcl}-E_J(1-\cos\varphi)}}{2\sqrt{E_C}} = \pi\left(n+\frac{1}{2}\right),
\nonumber\\
\label{BS0}
\end{eqnarray}
where
$
\varphi_0(E_n^{\rm qcl}) = \arccos\left( 1-{E_n^{\rm qcl}}/{E_J} \right)
$
are the classical turning points in the Josephson potential.
Evaluating the integral in  Eq. (\ref{BS0}), one arrives at the equation for $E_n^{\rm qcl}$ in the form
\begin{eqnarray}
\left( \frac{E_n^{\rm qcl}}{2E_J}-1 \right){\rm K}\left(\sqrt{\frac{E_n^{\rm qcl}}{2E_J}}\right) +  {\rm E}\left(\sqrt{\frac{E_n^{\rm qcl}}{2E_J}}\right)
\nonumber\\
= \frac{\pi}{2}\left(n+\frac{1}{2}\right)\sqrt{\frac{E_C}{2E_J}},
\label{BS1}
\end{eqnarray}
where ${\rm K}(k)$ and ${\rm E}(k)$ are the complete elliptic integrals of the first and the second kind (here, we use the definitions of these functions given in the book \cite{GR},
which differ from the definitions used in the popular Mathematica package),
and $n$ is the non-negative integer number
taking the values $0,1,2,\dots$.
Equation (\ref{BS1}) has solutions provided $E_J/E_C>\pi^2/32$, while for lower values of the ratio  $E_J/E_C$ no discrete levels exists in the potential well.
The total number of the energy levels in the well equals to $n_{\max}+1$, where $n_{\max}$ is given by
\begin{eqnarray}
n_{\max} =\left\lfloor \frac{1}{\pi}\sqrt{\frac{8E_J}{E_C}}-\frac{1}{2} \right\rfloor.
\label{nmax}
\end{eqnarray}
Here the brackets $\lfloor \dots \rfloor$ imply the floor function.
Having found the quasiclassical energy levels $E_n^{\rm qcl}$ from  Eq. (\ref{BS1}), we correct them as follows 
\begin{eqnarray}
E_n = E_n^{\rm qcl} - {E_C}/{8}.
\label{BS}
\end{eqnarray} 
We have numerically verified that the accuracy of the approximation (\ref{BS})
is 4\% for $E_J/E_C=3$,  0.3\% for $E_J/E_C=50$, and even better than that for higher values of the ratio $E_J/E_C$.
The maximum error occurs for the top most level $E_{n_{\max}}$, while the positions of the low lying levels are very accurate.
Next, we expand Eq. (\ref{BS1}) at small energies $E_n^{\rm qcl}\ll 2E_J$ and keep only the two leading terms in the small parameter $E_C/E_J$. 
In this way, we obtain an approximation for the low lying energy levels,
\begin{eqnarray}
E_n = \sqrt{8E_JE_C}\left(n+\frac{1}{2}\right) - \frac{E_C}{2}\left[\left(n+\frac{1}{2}\right)^2 + \frac{1}{4}\right].
\label{En}
\end{eqnarray}
It agrees with the well known result of  perturbation theory in weak anharmonicity of the Josephson potential \cite{Koch}.   

WKB approximation also allows one to find the half-bandwidths $\delta_n$. For the lowest band with $n=0$, 
the result is well known \cite{Averin,SZ},
\begin{eqnarray}
\delta_0 = -16\sqrt{\frac{E_JE_C}{\pi}}\left(\frac{E_J}{2E_C}\right)^{1/4}\exp\left[-\sqrt{\frac{8E_J}{E_C}}\right].
\end{eqnarray}
In order to find $\delta_n$ for $1\leq n \leq n_{\max}$, we use the WKB formula for the level splitting in
a double-well potential \cite{Grag}, 
\begin{eqnarray}
&&\delta_n = \frac{\hbar\omega_n}{\sqrt{\pi}}\times\,
\nonumber\\ && 
\exp\left[ - \frac{1}{2} - \int_{\varphi_0(E_n)}^{2\pi-\varphi_0(E_n)} d\varphi\frac{\sqrt{E_J(1-\cos\varphi)-E_n}}{2\sqrt{E_C}} \right].
\label{dn}
\end{eqnarray}
Here, we have introduced the frequency of classical oscillations at the bottom the potential well for a particle with the energy $E_n$,
\begin{eqnarray}
\omega_n = \frac{\pi\sqrt{2E_JE_C}}{\hbar {\rm K}\left(\sqrt{{E_n}/{2E_J}}\right)}.
\label{omega_n}
\end{eqnarray}
Evaluating the integral in  Eq. (\ref{dn}), we arrive at
\begin{eqnarray}
&& \delta_n = \frac{(-1)^{n+1}\sqrt{2\pi E_JE_C}}{{\rm K}\left(\sqrt{{E_n}/{2E_J}}\right) }
\exp\left\{- \frac{1}{2} - \sqrt{\frac{8E_J}{E_C}} 
\right.
\nonumber\\  &&
\left. 
\times\,\left[ \,{\rm E}\left(\sqrt{1-\frac{E_n}{2E_J}}\right)
-\frac{E_n}{2E_J}\,{\rm K}\left(\sqrt{1-\frac{E_n}{2E_J}}\right) \right] \right\}.
\nonumber\\
\label{delta_n}
\end{eqnarray}
Comparison with the exact numerical simulation shows that the approximation (\ref{delta_n}) is quite accurate.
It has the accuracy 10\% or better for all bands with energies below the barrier top and for ratios
$E_J/E_C>3$. The maximum relative error in $\delta_n$ again occurs
for the highest Bloch band inside the potential well, namely for $\delta_{n_{\max}}$.

Now, we turn to the matrix elements of the phase operator (\ref{phi_mn}).
We use the well known quasiclassical approximation for the matrix elements \cite{LL,Nikitin},
\begin{eqnarray}
|\varphi_{mn}| &\approx& \frac{\omega(\varepsilon_{mn}^+)}{2\pi}\int_{-{\pi}/{\omega(\varepsilon_{mn}^+)}}^{{\pi}/{\omega(\varepsilon_{mn}^+)}} dt 
\,\varphi_{\rm cl}\left(t,\varepsilon_{mn}^+\right)
\nonumber\\ &&\times\,
\sin\big[|m-n|\omega(\varepsilon_{mn}^+)t\big].
\label{ME1}
\end{eqnarray}
Here $\varepsilon_{mn}^+=(E_n^{\rm qcl}+E_m^{\rm qcl})/2$, the frequency
$\omega(\varepsilon_{mn}^+)$ is given by Eq. (\ref{omega_n}) with $E_n$ replaced by $\varepsilon_{mn}^+$, and
\begin{eqnarray}
\varphi_{\rm cl}(t,\varepsilon_{mn}^+) = 2\,{\rm am}\left( 2\sqrt{E_C\varepsilon_{mn}^+}\,t,\, \sqrt{\frac{2E_J}{\varepsilon_{mn}^+}} \right) 
\end{eqnarray}
is the solution of the classical equation of motion for a particle in the Josephson potential well having the energy $\varepsilon_{mn}^+$. 
It is expressed in terms of Jacobi amplitude function ${\rm am}(u,k)$. The integral (\ref{ME1}) can be solved
analytically using the properties of Jacobi functions\cite{GR}. We find
$\varphi_{mn}=0$ if $|m-n|$ is an even number, and 
\begin{eqnarray}
|\varphi_{mn}|=\frac{4}{|m-n|} \frac{q_{mn}^{|m-n|/2}}{1+q_{mn}^{|m-n|}}
\label{phi_mn}
\end{eqnarray}
if $|m-n|$ is an odd number. The parameter $q_{mn}$ is known as the so called nome in the theory of Jacobi functions,
\begin{eqnarray}
q_{mn}=\exp\left[ -\pi \frac{{\rm K}\left(\sqrt{1-{\varepsilon_{mn}}/{2E_J}}\right)}{{\rm K}\left(\sqrt{{\varepsilon_{mn}}/{2E_J}}\right)} \right].
\end{eqnarray} 
For completeness, we also provide the quasiclassical matrix elements of the momentum (\ref{pmn}) for the odd $|m-n|$ (for even $|m-n|\not=0$ they are also equal to zero),
\begin{eqnarray}
|p_{mn}|=\left|\left\langle m\left| -i\frac{\partial}{\partial\varphi} \right| n\right\rangle\right| = \frac{|E_m-E_n||\varphi_{mn}|}{8E_C}
\nonumber\\
= \frac{|E_m-E_n|}{2E_C|m-n|} \frac{q_{mn}^{|m-n|/2}}{1+q_{mn}^{|m-n|}}.
\label{p_mn}
\end{eqnarray}

For the low lying levels with $E_n\ll 2E_J$ one can derive simpler expressions.
In this limit, one can use an approximation 
$E_J(1-\cos\varphi)\approx E_J(\varphi^2/2 - \varphi^4/24)$ and treat the term $\propto\varphi^4$ as a perturbation. Keeping the two lowest order terms, we find
\begin{eqnarray}
\varphi_{n-1,n} = \varphi_{n,n-1}
= \left(\frac{2E_C}{E_J}\right)^{1/4}\sqrt{n}\left[ 1+n\sqrt{\frac{E_C}{32E_J}} \right].
\nonumber\\
\label{phinm}
\end{eqnarray}
For comparison, we also expand the quasiclassical matrix element (\ref{phi_mn}) in the same limit and find
\begin{eqnarray}
|\varphi_{n,n-1}|
= \left(\frac{2E_C}{E_J}\right)^{1/4}\sqrt{n}\left[ 1+\sqrt{\frac{E_C}{32E_J}}\left(n-\frac{1}{8n}\right) \right].
\nonumber\\
\end{eqnarray}
The two expressions agree quite well even for $n=1$, which confirms the accuracy of the approximation (\ref{phi_mn}). 
Numerics shows that the approximate expression for the matrix element of momentum $p_{n,n-1}$, given by  Eq. (\ref{p_mn}),
is very accurate. Namely, for $E_J/E_C=3$, we find a tiny error of 0.3\%, while for $E_J/E_C=50$ the maximum error, occurring at $n=n_{\max}=5$, turns out to be 2\%.    

The  matrix element between the states $n$ and $n\pm 3$, derived by perturbation theory in the quartic term, reads
\begin{eqnarray}
\varphi_{n-3,n} = \varphi_{n,n-3}
= -\left(\frac{2E_C}{E_J}\right)^{3/4}\frac{\sqrt{n(n-1)(n-2)}}{48}.
\nonumber\\
\label{phinm3}
\end{eqnarray}
Expanding the quasiclassical expression (\ref{phi_mn}) in powers of the small parameter $E_C/E_J$,  we find
\begin{eqnarray}
|\varphi_{n-3,n}| = |\varphi_{n,n-3}| = \left(\frac{2E_C}{E_J}\right)^{3/4}\frac{(n-1)^{3/2}}{48}.
\label{phi_nm3}
\end{eqnarray}
Again, we note that the two expressions (\ref{phinm3}) and (\ref{phi_nm3}) agree quite well. 
Numerically, we find that for $E_J/E_C=50$, in which case $n_{\max}=5$, 
the maximum relative error for the matrix element $p_{n,n-3}$, given by  Eq. (\ref{p_mn}), equals to 7\% and occurs at $n=3$.

\section{Solution of the master equation for a harmonic oscillator}
\label{Harmonic_app}

In this Appendix, we find the stationary solution of the master equation (\ref{master}) 
in harmonic approximation, and use it to derive the expression  (\ref{P1_osc_1}) for the heat flux.
If one replaces the nonlinear Josephson potential by a harmonic one, $E_J(1-\cos\varphi)\to E_J\varphi^2/2$, 
the energy levels become equidistant, $E_n=\hbar\omega_J(n+1/2)$, and independent of $q$.
In this approximation,  one finds $\omega_{n+1,n}=\omega_J$ for all $n$. 
The phase matrix elements connecting neighboring levels are given by
\begin{eqnarray}
\varphi_{n-1,n}=\varphi_{n,n-1}=\left(\frac{2E_C}{E_J}\right)^{1/4}\sqrt{n}, \;\;\; n=1,2,3,\dots
\end{eqnarray}
All other matrix elements vanish. Accordingly, the transition rates between the levels (\ref{Gamma_mn}) take the form
\begin{eqnarray}
\Gamma_{n,n-1}^{(j)} = n \gamma_\uparrow^{(j)},\;\;\; \Gamma_{n-1,n}^{(j)} = n \gamma_\downarrow^{(j)}
\end{eqnarray}
where we have defined
\begin{eqnarray}
\gamma_{\uparrow}^{(j)} &=& \frac{2E_C}{e^2}\,{\rm Re}\left[\frac{1}{Z_j(\omega_J)}\right] N_j(\omega_J),
\nonumber\\
\gamma_{\downarrow}^{(j)} &=& \frac{2E_C}{e^2}\,{\rm Re}\left[\frac{1}{Z_j(\omega_J)}\right][1+N_j(\omega_J)].
\label{gamma}
\end{eqnarray}
In this approximation, the solution
of Eq. (\ref{master}) can be found analytically,
\begin{eqnarray}
W_m = u^m (1-u),\; u= (\gamma_\uparrow^{(1)}+\gamma_\uparrow^{(2)})/(\gamma_\downarrow^{(1)}+\gamma_\downarrow^{(2)}).
\label{harmonic_probabilities}
\end{eqnarray}
Substituting this result in the general expression for the heat flux (\ref{Q1_weak}), one arrives at the 
formula (\ref{P1_osc_1}). 

If we allow anharmonicity, but consider only the transitions between the neighboring
levels, the occupation probabilities for an N-level system are 
\begin{eqnarray}
W_0 = \frac{1}{N_c},\;\;  W_m = \frac{1}{N_c}\prod_{p=1}^m\frac{\Gamma_{p,p-1}}{\Gamma_{p-1,p}},
\label{anharmonic_probabilities}
\end{eqnarray}
where $ m=1,2,...,N-1$, and
$$
N_c=1+\sum_{m=1}^{N-1}\prod_{p=1}^m\frac{\Gamma_{p,p-1}}{\Gamma_{p-1,p}},\;\; 
\Gamma_{k,l}=\Gamma_{k,l}^{(1)}+\Gamma_{k,l}^{(2)}
$$
As $N\rightarrow\infty$ and with vanishing anharmonicity, (\ref{anharmonic_probabilities}) reduces to (\ref{harmonic_probabilities}).

\section{Derivation of  Eq. (\ref{Pb_final})}
\label{charge}

In this appendix, we provide the details of the derivation of the expression (\ref{Pb_final})
for the intraband contribution to the heat flux. We ignore the interband transitions and put $\delta_n=0$.
In this approximation, the Hamiltonian of the junction, $\hat H_J$, drops out from the full Hamiltonian (\ref{H_full}). 
The operator of the Josephson phase can be expressed as $\hat\varphi = -2ie (\partial / \partial q)$, 
hence the interaction Hamiltonians may be written in the form
\begin{eqnarray}
\hat H_j^{\rm int} = \sum_k \left( 2i e c_{j,k}\hat X_{j,k}\frac{\partial}{\partial q} - \frac{2e^2c_{j,k}^2}{M_{j,k}\omega_{j,k}^2}\frac{\partial^2}{\partial q^2} \right). 
\label{Hint_q}
\end{eqnarray}
Since after making these approximations 
the Hamiltonian has become quadratic in all operators, one can use Wick's theorem 
and the Baker-Campbell-Hausdorff formula for the commutators. 
This leads to 
\begin{eqnarray}
&& \left\langle e^{i\pi\hat q(t)/e} e^{-i\pi\hat q(0)/e} \right\rangle 
= \left\langle e^{i\frac{\pi}{e}(\hat q(t)-\hat q(0))}  \right\rangle e^{\frac{\pi^2}{2e^2}[\hat q(t),\hat q(0)]}
\nonumber\\ &&
=\, e^{-\frac{\pi^2}{2e^2}\langle(\hat q(t)-\hat q(0))^2\rangle} e^{\frac{\pi^2}{2e^2}[\hat q(t),\hat q(0)]}.
\end{eqnarray}
It is straightforward to show by solving equations of motion for the quantum operators,
that for the Hamiltonian of the form $\hat H_1+\hat H_2 + \hat H_1^{\rm int} + \hat H_2^{\rm int}$, with the interaction terms given
by  Eq. (\ref{Hint_q}), the charge correlators read 
\begin{eqnarray}
\frac{\pi^2}{2e^2}\langle(\hat q(t)-\hat q(0))^2\rangle &=& {\cal F}_1(t) + {\cal F}_2(t),
\nonumber\\ 
\frac{\pi^2}{2e^2}[\hat q(t),\hat q(0)] &=& -i{\cal K}_1(t) - i{\cal K}_2(t),
\end{eqnarray}
where the functions ${\cal F}_j(t)$ and ${\cal K}_j(t)$  are given by the integrals (\ref{calF}) and (\ref{calK}).
Now one can find the probabilities ${\cal P}_j(\omega)$ from  Eq. (\ref{Pj_q}), and ${\cal P}(\omega)$  from  Eq. (\ref{convolution_1}).

Next, we substitute the correlation function (\ref{SQ_b}) in  Eq. (\ref{Q3}) and find  the intraband contribution to the heat flux in the form
\begin{eqnarray}
J^{\rm b} &=&  \left(\sum_{n=0}^{n_{\max}} \frac{\pi^3\delta_n^2}{e^2} W_n\right) 
\int\frac{d\omega}{2\pi}  \,{\rm Re}\left[\frac{1}{Z_1(\omega)}\right] 
\nonumber\\ && \times\,
\big({\cal P}(-\omega)[1+N_1(\omega)]-{\cal P}(\omega)N_1(\omega)\big).
\label{P1_b_general}
\end{eqnarray}
We can further transform $J^{\rm b}$ by applying a useful property of the function ${\cal P}_1(\omega)$,
which is derived by applying the Fourier transformation to both sides of the identity
\begin{eqnarray}
\frac{d}{dt}e^{-{\cal F}(t)-i{\cal K}(t)} = -\left(\dot{\cal F}(t)+i\dot{\cal K}(t)\right)e^{-{\cal F}(t)-i{\cal K}(t)}.
\end{eqnarray}
Since the functions ${\cal F}$ and ${\cal K}$ are given by the integrals (\ref{calF}) and (\ref{calK}), we arrive at the result
\begin{eqnarray}
\int d\omega' \,{\rm Re}\left[\frac{1+N_1(\omega')}{Z_1(\omega')}\right]  {\cal P}_1(\omega-\omega')=\frac{e^2\omega}{\pi\hbar} {\cal P}_1(\omega).
\label{proper}
\end{eqnarray}
With the aid of this identity one can easily transform the heat flux (\ref{P1_b_general}) to the form (\ref{Pb_final}).

\section{Derivation of  Eq. (\ref{Pin})}
\label{Sphi}

In this appendix, we derive the expression  (\ref{phi_phi}) for the phase-phase correlation function by means of perturbation theory in small Josephson energy $E_J$.
Considering the term $-E_J\cos\varphi$ in the full Hamiltonian (\ref{H_full}) as a perturbation 
and keeping the terms up to $E_J^2$ in the expansion, one can express the 
time dependent phase operator $\hat\varphi(t)=e^{i\hat Ht/\hbar} \hat\varphi e^{-i\hat Ht/\hbar}$ in the form
\begin{eqnarray}
&&\hat\varphi(t) = \hat\varphi_0(t) - \frac{iE_J}{\hbar}\int_0^t dt' \big[ \cos\hat\varphi_0(t'), \hat\varphi_0(t) \big]
\nonumber\\ &&
-\, \frac{E_J^2}{\hbar^2}\int_0^t dt'\int_0^{t'} dt''\big[ \cos\hat\varphi_0(t''),\big[ \cos\hat\varphi_0(t'),\hat\varphi_0(t) \big] \big].
\nonumber\\ \label{phi_t}
\end{eqnarray} 
Here the time evolution of the operator $\hat\varphi_0(t) = e^{i\hat H_0t/\hbar} \hat\varphi e^{-i\hat H_0t/\hbar}$
is determined by quadratic Hamiltonian $\hat H_0 = \lim_{E_J\to 0} \hat H$. As in the previous appendix, this ensures the validity of the Wick's theorem
for various products involving phase operators. In particular, with the aid of the Wick's theorem one can prove that
\begin{eqnarray}
\big[ \cos\hat\varphi_0(t'), \hat\varphi_0(t) \big] = \sin\hat\varphi_0(t') \big[\hat\varphi_0(t),\hat\varphi_0(t')\big],
\label{comm1}
\end{eqnarray}
\begin{eqnarray}
&& \big[ \cos\hat\varphi_0(t''),\big[ \cos\hat\varphi_0(t'),\hat\varphi_0(t) \big] \big]
\nonumber\\ &&
=\, \big( \sin[\hat\varphi_0(t')+\hat\varphi_0(t'')] - \sin[\hat\varphi_0(t')-\hat\varphi_0(t'')] \big)
\nonumber\\ &&\times\,
\sinh\left(\frac{\big[\hat\varphi_0(t'),\hat\varphi_0(t'')\big]}{2}\right)\big[\hat\varphi_0(t),\hat\varphi_0(t')\big].
\label{comm2}
\end{eqnarray}
Here, we have also used the property typical for quadratic Hamiltonians, namely, we have used the fact that
the commutator $\big[\hat\varphi_0(t),\hat\varphi_0(t')\big]$ is proportional to the identity operator $\hat E$ 
and commutes with all other operators,
\begin{eqnarray}
\big[\hat\varphi_0(t),\hat\varphi_0(t')\big] = -2i K(t-t')\hat E.
\end{eqnarray} 
From the equation of motion for $\hat\varphi_0(t)$,
or applying path integral techniques \cite{Grabert},  
one finds $K(t)=K_1(t)+K_2(t)$, with the functions $K_{1,2}(t)$ having the form (\ref{FKK}). 

Using the expansion (\ref{phi_t}), transforming the commutators in it according to the rules (\ref{comm1}) and (\ref{comm2}),
and taking the long-time limit, we express the phase-phase correlation function in the form 
\begin{eqnarray}
&& \langle\hat\varphi(t_1)\hat\varphi(t_2)\rangle =  \langle\hat\varphi_0(t_1)\hat\varphi_0(t_2)\rangle 
\nonumber\\ &&
+\, \frac{I_C^2}{e^2} \int dt' dt'' K(t_1-t')K(t_2-t'')\langle\sin\hat\varphi_0(t')\,\sin\hat\varphi_0(t'')\rangle
\nonumber\\ &&
-\,\frac{I_C^2}{2e^2}\int dt'\int_{-\infty}^{t'} dt'' K(t_2-t')\sin[K(t'-t'')]
\nonumber\\ && \times\,
\langle \hat\varphi_0(t_1) \sin[\hat\varphi_0(t')-\hat\varphi_0(t'')] \rangle
\nonumber\\ &&
-\, \frac{I_C^2}{2e^2}\int dt'\int_{-\infty}^{t'} dt'' K(t_1-t')\sin[K(t'-t'')]
\nonumber\\ && \times\,
\langle  \sin[\hat\varphi_0(t')-\hat\varphi_0(t'')]\hat\varphi_0(t_2)\rangle.
\label{phi_phi_1}
\end{eqnarray}
Here, we have omitted the terms containing $\sin[\hat\varphi_0(t')+\hat\varphi_0(t'')]$, which
vanish upon averaging because at $E_J=0$ the phase fluctuations are unrestricted. 
Next, we apply Wick's theorem once again and find
\begin{eqnarray}
&& \langle \hat\varphi_0(t_1) \sin[\hat\varphi_0(t')-\hat\varphi_0(t'')] \rangle
\nonumber\\ &&
=\, \langle \hat\varphi_0(t_1)[\hat\varphi_0(t')-\hat\varphi_0(t'')] \rangle
\langle \cos[\hat\varphi_0(t')-\hat\varphi_0(t'')] \rangle
\nonumber\\ &&
=\, e^{-F(t'-t'')}\langle \hat\varphi_0(t_1)[\hat\varphi_0(t')-\hat\varphi_0(t'')] \rangle,
\end{eqnarray}
where $F(t)=F_1(t)+F_2(t)$ and the functions $F_{1,2}(t)$ are given by Eqs. (\ref{FKK}).
Similarly,
\begin{eqnarray}
&& \langle  \sin[\hat\varphi_0(t')-\hat\varphi_0(t'')]\hat\varphi_0(t_2)\rangle
\nonumber\\ &&
=\, e^{-F(t'-t'')}\langle [\hat\varphi_0(t')-\hat\varphi_0(t'')]\hat\varphi_0(t_2)\rangle.
\end{eqnarray}
Substituting these expressions in  Eq. (\ref{phi_phi_1}), and taking the Fourier transformation over the time difference $t_1-t_2$, 
we find
\begin{eqnarray}
S_\varphi(\omega) &=& \left( 1 + i \frac{\hbar\omega K_\omega}{2e^2 Z_J(\omega)}  -  i \frac{\hbar\omega K^*_\omega}{2e^2 Z_J^*(\omega)} \right)S_{\varphi}^{(0)}(\omega)
\nonumber\\ &&
+\, \frac{I_C^2}{e^2}|K_\omega|^2 S_{\sin\varphi}(\omega).
\label{phi_phi_app} 
\end{eqnarray}
The junction impedance $Z_J(\omega)$, appearing here, is defined by  Eq. (\ref{ZJ1}), the function $K_\omega$ reads
\begin{eqnarray}
K_\omega=\int_0^\infty e^{i\omega t} K(t)
= \frac{2e^2}{-i\hbar\omega} \frac{1}{-i\omega C + \frac{1}{Z_1(\omega)}+\frac{1}{Z_2(\omega)} },
\nonumber\\
\end{eqnarray}
the Fourier transformed phase-phase correlation function evaluated at $E_J=0$ is given by
\begin{eqnarray}
&& S_\varphi^{(0)}(\omega) = \int dt e^{i\omega t} \langle\hat\varphi_0(t)\hat\varphi_0(0)\rangle
\nonumber\\ &&
=\,\frac{2\hbar\omega}{e^2}|K_\omega|^2\,{\rm Re}\left[\frac{1+N_1(\omega)}{Z_1(\omega)} + \frac{1+N_2(\omega)}{Z_2(\omega)}\right],
\end{eqnarray}
and the correlation function of $\sin\hat\varphi_0$ is defined as
\begin{eqnarray}
S_{\sin\varphi}(\omega) = \int dt e^{i\omega t} \langle\sin\hat\varphi_0(t)\,\sin\hat\varphi_0(0)\rangle.
\end{eqnarray}
One can straightforwardly show that $S_{\sin\varphi}(\omega)=\pi P(\omega)$, where $P(\omega)$ is the photon emission probability (\ref{P}). 
Comparing  Eqs. (\ref{phi_phi_app}) and (\ref{phi_phi}) one can verify that they coincide in the lowest nonvanishing order
of the perturbation theory $\sim E_J^2$, but Eq. (\ref{phi_phi}) has more compact and physically transparent form. 

The inelastic contribution to the heat flux originates from the last term in the correlation function (\ref{phi_phi_app})
containing $S_{\sin\varphi}(\omega)$. Substituting this term in the general expression (\ref{Q30}), we obtain
\begin{eqnarray}
J_{1}^{\rm inel} = I_C^2 \int d\omega {\cal R}_1(\omega)[1+N_1(\omega)]P(-\omega).
\label{Pin0}
\end{eqnarray}
One can transform this expression to a more physically meaningful form (\ref{Pin}) invoking the property analogous to (\ref{proper}),
\begin{eqnarray}
\int d\omega'\,  {\cal R}_1(\omega') [1+N_1(\omega')]  P_1(\omega-\omega') = \frac{\pi\hbar \omega}{4e^2}   P_1(\omega).
\nonumber\\
\end{eqnarray}

\end{document}